\documentclass[aps,pra,10pt,showpacs,twocolumn,superscriptaddress,longbibliography]{revtex4-1}
\usepackage{graphicx}
\usepackage[usenames]{color}
\usepackage{amssymb,amsmath}
\usepackage{siunitx}
\usepackage{xcolor}
\usepackage{setspace} 
\usepackage{float} 
\usepackage{tabularx}
\usepackage[linkcolor=blue,citecolor=blue,colorlinks=true,urlcolor=blue]{hyperref}
\newcommand{\eq}[1]{Eq.~(\ref{#1})}
\newcommand{\fig}[1]{Fig.~\ref{#1}}
\newcommand{\be}[1]{\begin{equation}\label{#1}}
\newcommand{\ee}{\end{equation}}
\usepackage{amsmath,esint}
\usepackage{comment}
\usepackage{lipsum}
\usepackage[toc]{appendix}
\begin{document}

\title{A general model and toolkit for the ionization of three or more electrons in strongly driven molecules using an effective Coulomb potential for the interaction between bound electrons }

\author{G. P. Katsoulis}
\affiliation{Department of Physics and Astronomy, University College London, Gower Street, London WC1E 6BT, United Kingdom}
\author{M. B. Peters}
\affiliation{Department of Physics and Astronomy, University College London, Gower Street, London WC1E 6BT, United Kingdom}
\author{A. Emmanouilidou}
\affiliation{Department of Physics and Astronomy, University College London, Gower Street, London WC1E 6BT, United Kingdom}
\begin{abstract}

We formulate a general three-dimensional semiclassical model   for the study of correlated multielectron escape during fragmentation of molecules driven by intense infrared laser pulses, while fully accounting for the magnetic field of the laser pulse. We do so in the context of triple ionization of strongly driven HeH$_{2}^{+}$. 
 Our model fully accounts for the singularity in the Coulomb potentials of  a recolliding electron with the core and a bound electron with the core as well as for 
the interaction of a recolliding with a bound electron. To avoid artificial autoionization, our model  employs effective potentials to treat  the interaction between bound electrons.
  We focus on triple and double ionization as well as frustrated triple and frustrated double ionization. In these processes, we identify and explain the main features of the sum of the kinetic energies of the final ion fragments.
 We find that frustrated double ionization is a major ionization process, and we identify the different channels and hence different final fragments that are obtained through frustrated double ionization. Also, we discuss the differences 
 between frustrated double and triple ionization. 
 
 \end{abstract}

\date{\today}

\maketitle

\section{Introduction}

Multielectron ionization and  the formation of highly excited Rydberg states  are fundamental processes in molecules driven by intense and infrared laser pulses. Rydberg states  have a wide range of applications, such as, acceleration of neutral particles \cite{Eichmann}, spectral features of photoelectrons \cite{Veltheim}, and formation of molecules via long-range interactions \cite{Bendkowsky}. 
Electron-nuclear correlated multi-photon resonance excitation was shown to be the mechanism responsible for the formation of Rydberg states in weakly-driven H$_{2}$ \cite{Zhang}. This process was shown to merge  with frustrated double ionization for H$_{2}$ driven by intense infrared laser fields---strongly driven \cite{Manschwetus,Emmanouilidou_2012}.   In frustrated ionization an electron first tunnel ionizes in the driving laser field. Then, due to the electric field, this electron is recaptured by the parent ion in a Rydberg state \cite{Nubbemeyer}.

Most studies on strongly driven molecules address double and frustrated double ionization  of  two-electron molecules \cite{Emmanouilidou_2012,PhysRevA.94.043408,Agapi_2016,PhysRevA.96.033404,PhysRevA.101.033403,PhysRevA.108.013120,Zhang2023,Manschwetus,Zhang2,McKenna1, MScKenna2, Sayler,Tiwari2022}. However, there are scarcely any theoretical studies on three-electron ionization and formation of Rydberg states during the fragmentation of strongly driven molecules. The reason is that accounting for both multielectron and nuclear motion is a formidable task. This is corroborated by the few theoretical studies  for   three-electron ionization of strongly driven atoms, a simpler but still highly challenging task. Mostly formulated in the dipole approximation, these studies on atoms employ lower dimensionality classical \cite{PhysRevA.64.053401,PhysRevLett.97.083001} and quantum mechanical \cite{PhysRevA.98.031401,Efimov:21} models to reduce the complexity and computational resources required. However, lower dimensionality results in a non accurate description of electron-electron interaction during triple ionization. Currently, only classical or semiclassical 3D models of triple ionization of atoms are available \cite{PhysRevLett.97.083001,Zhou:10,Tang:13,PhysRevA.104.023113,PhysRevA.105.053119}.

 In Refs. \cite{Agapi3electron,PhysRevA.107.L041101}, we argued that the main disadvantage of available classical and quantum models of triple ionization in atoms is their softening of the Coulomb interaction of each electron with the core.   In quantum mechanical models, this interaction is softened to obtain a  computationally tractable problem.
However,  in classical and semiclassical models, softening the Coulomb singularity is fundamental and relates to unphysical autoionization. Classically there is no lower energy bound. Hence, when 
a bound electron undergoes a close encounter with the core, the Coulomb singularity  allows this electron to acquire a very negative energy. Through the Coulomb interaction, this energy can be shared  by another bound electron potentially leading to its artificial escape.

To avoid artificial autoionization, most classical and semiclassical models of triple ionization of strongly driven atoms soften the Coulomb potential \cite{PhysRevLett.97.083001,Zhou:10,Tang:13} or add Heisenberg potentials  \cite{PhysRevA.21.834} (effective softening) to mimic the Heisenberg uncertainty principle and prevent each electron from a close encounter with the core \cite{PhysRevA.104.023113,PhysRevA.105.053119}.
However, softening the Coulomb potential does not allow for an accurate  description of  electron scattering from the core \cite{Pandit2018,Pandit2017}.
This  is due to the exponential decrease  of the ratio of the scattering amplitude for the soft-core potential over the one for the Coulomb potential  with increasing momentum transfer \cite{Pandit2018,Pandit2017}. For recollisions \cite{Corkum_1994}, this implies that soft potentials are quite inaccurate for high energy recolliding electrons that backscatter. 
 Indeed, we have shown \cite{Agapi3electron,PhysRevA.107.L041101} that the ionization spectra obtained with  the Heseinberg model     differ from experimental ones  obtained for driven Ne and Ar\cite{PhysRevLett.84.447,Rudenko_2008,PhysRevA.86.043402,Herrwerth_2008,Zrost_2006,PhysRevLett.93.253001,SHIMADA2005221}.

  Here, we formulate a general 3D semiclassical model of multielectron ionization and fragmentation  of strongly driven molecules, while  fully accounting for the magnetic field of the laser pulse.  The main premise in our model is that two interactions are most important during a recollision or return of an electron to the core and hence are treated exactly. We account for the singularity in the Coulomb potential between each electron, bound or quasifree, and the core. Quasifree refers to a recolliding electron or an electron escaping to the continuum. Also, we treat exactly the Coulomb potential between each pair of a quasifree and a bound electron and hence the transfer of energy from a quasifree to a bound electron. However, accounting for  the  singularity in the electron-core interaction, implies we need  to avoid  unphysical autoionization that can take place through energy transfer between bound electrons. We do so by 
  approximating the energy transfer from a bound to a bound electron via the use 
  of effective Coulomb potentials. We refer to this model that accurately accounts for all Coulomb interactions and only employs approximate effective Coulomb potentials to describe the bound-bound electron interaction as ECBB model. This is a sophisticated model that identifies during time propagation whether an electron is quasifree or bound. That is, in the ECBB model, we decide on the fly if the full or effective Coulomb potential  describes the interaction between a pair of electrons. We do so by using a set of simple criteria detailed below.
  
 The ECBB model for strongly driven molecules is a generalization of the ECBB model we have previously developed for strongly driven atoms \cite{Agapi3electron,PhysRevA.107.L041101}. For atoms, we have shown that the ECBB model  results in triple ionization spectra of strongly driven Ne that are in excellent agreement with experiment \cite{PhysRevA.107.L041101}. The main difficulty in generalizing the ECBB model  to molecules is the formulation of the effective Coulomb potential between a pair of bound electrons in the presence of many nuclei. In atoms, the effective Coulomb potential is obtained by assuming that a  bound electron  has a charge distribution  centered around the atomic core. In molecules, we assume that an electron creates an effective potential that is the sum of the effective Coulomb potentials with respect to each nucleus separately, i.e., each effective Coulomb potential is the same as  for an atom. However, each atomic effective Coulomb potential is weighted by an approximate expression for the probability to find the  electron that creates this potential at a given position around this nucleus.

 Using the ECBB model, we address triple and double ionization in the  triatomic molecule HeH$_{2}^{+}$ when driven by a linearly polarized laser pulse. We also address the formation of Rydberg states via frustrated ionization. 
 Employing the HeH$_{2}^{+}$ molecule allows us to directly compare the results for triple, double and frustrated triple ionization obtained with the ECBB model  in this study and with a predecessor of the ECBB model  obtained in previous work  \cite{PhysRevA.103.043109}.  This previous model determined on the fly whether an electron is quasifree or bound, as the ECBB model does, but  this was done using a set of criteria less sophisticated than the ones employed in the ECBB model for atoms \cite{Agapi3electron,PhysRevA.107.L041101} and molecules in the current work. Also, in this previously  employed  model    the interaction of a pair of bound electrons was set equal to zero.  As a result, in Ref. \cite{PhysRevA.103.043109}, we could only address frustrated triple ionization where two electrons ionize and one electron remains bound in a Rydberg state. Here, since we account via effective potentials for the interaction between bound electrons, we can also address  frustrated double ionization. In this process,  one electron ionizes, the other remains bound in the ground state of one of the He or H fragments and another electron remains bound  in a Rydberg state of one of the fragments.

Here, we address both triple and double ionization (TI and DI) as well as frustrated triple and double ionization (FTI and FDI) and obtain the sum of the kinetic energies of the atomic fragments, referred to as kinetic energy release. We find that the atomic fragments are moving faster when fragmentation of  HeH$_{2}^{+}$  is described with the ECBB model compared to its predecessor \cite{Agapi3electron,PhysRevA.107.L041101}. This is consistent with our finding that the electron-electron escape to the continuum is more correlated when the process is described with the ECBB model, since it allows for energy transfer between bound electrons. Higher correlation results in faster electron escape  leading to  Coulomb explosion of the nuclei at smaller distances, eventually resulting in higher kinetic energies.
 Also, we find that,  in three-electron molecules,  FTI and FDI  proceed via two pathways, first identified in FDI of the strongly driven two-electron molecule H$_{2}$ \cite{Emmanouilidou_2012}. One electron ionizes early on (first step), while the remaining bound electron does so later in time (second step). If the second (first) ionization step is frustrated, we label the FTI and FDI pathway as A and B, respectively.

\section{Method}\label{Sec::Method}
In what follows, we describe in detail the formulation of the ECBB model for strongly driven molecules. The ECBB model resolves unphysical autoionization in 3D semiclassical models that fully account for  the Coulomb singularity. Also, we formulate the ECBB model in the nondipole approximation fully accounting for the magnetic field component of the laser field. Finally, both electrons and the cores are propagated in time.
\subsection{Definition of the effective charge and potential}\label{Sec::MethodA}
In what follows, the cores are assigned indices in the interval $[1,N_{\text{c}}]$, where $N_{\text{c}}$ is the number of cores. The remaining indices in the interval $[N_{\text{c}}+1,N]$, with N  the number of particles, are assigned to the electrons.
For each electron $j$, we define an effective charge $\zeta_{j,n}(t)$ associated with each core $n$ as follows \cite{PhysRevA.40.6223}  
\begin{equation}\label{eqn::zeta_and_energy}
\zeta_{j,n}(t) = \left\{
    \begin{array}{ll}
        Q_n & \mathcal{E}_{j}(t)  \leq \mathcal{E}^{n}_{{1 s}}\\
        \left(Q_n / \mathcal{E}^{n}_{{1 s}}\right) \mathcal{E}_{j}(t) & \mathcal{E}^{n}_{{1 s}} < \mathcal{E}_{j}(t)  < 0\\
        0 & \mathcal{E}_{j}(t)  \geq 0,
    \end{array}
\right.
\end{equation}
where $Q_n$ is the charge of the core $n$, $\mathcal{E}^{n}_{{1 s}}$ is the ground-state energy of a hydrogenic atom with core charge $Q_n$, i.e., $\mathcal{E}^{n}_{{1 s}} = -Q^2_n/2,$ and  $\mathcal{E}_{j}(t)$ is the energy of electron $j$ given by 
\begin{align}\label{eq:energy_of_electron_j}
\begin{split}
{\mathcal{E}_{j}(t) }&= \frac{\left[\mathbf{\tilde{p}}_{j}- Q_j\mathbf{A}(\mathbf{r}_{j},t) \right]^2}{2m_j} + \sum_{n=1}^{N_{\text{c}}}\frac{Q_n Q_j}{|\mathbf{r}_{n}-\mathbf{r}_{j}|}  - Q_j\mathbf{r}_{j} \cdot \mathbf{E}\left(\mathbf{r}_{j}, t\right) \\
&+ \sum_{n=1}^{N_{\text{c}}}\sum_{\substack{\;{i=N_{\text{c}}+1} \\ {i} \neq {j}}}^{N} c_{i,j}(t)C_{i,n}(\mathcal{E}_{i},|\mathbf{r}_{1}-\mathbf{r}_{i}|,...,|\mathbf{r}_{N_{\text{c}}}-\mathbf{r}_{i}|) \times\\
&\times V_{\text{eff}}(\zeta_{i,n}(t),|\mathbf{r}_{n}-\mathbf{r}_{j}|),
\end{split}
\end{align}
where $\mathbf{A}(\mathbf{r}_{j},t)$ is the vector potential and $\mathbf{E}(\mathbf{r}_{j},t) = -\frac{\partial \mathbf{A}(\mathbf{r}_{j},t)}{\partial t} $ is the electric field. The effective potential that an electron $j$ experiences at a distance $|\mathbf{r}_{n}-\mathbf{r}_{j}|$ from the core $n$  due to the charge of electron $i$  is given by \cite{PhysRevA.40.6223}
\begin{align}\label{eq:eff_potential}
\begin{split}
&V_{\text{eff}}(\zeta_{i,n}(t),|\mathbf{r}_{n}-\mathbf{r}_{j}|) \\
& = \frac{1 - (1+\zeta_{i,n}|\mathbf{r}_{n}-\mathbf{r}_{j}|)e^{-2\zeta_{i,n}|\mathbf{r}_{n}-\mathbf{r}_{j}|}}{|\mathbf{r}_{n}-\mathbf{r}_{j}|}.
\end{split}
\end{align}

To generalize this effective Coulomb potential to molecules, in \eq{eq:energy_of_electron_j}, we assume  that an electron $i$ screens each core $n$ seperately as if it was an atom, with probability $|C_{i,n}|^2$ given by
\begin{align}\label{Cim_constant}
\begin{split}
C_{i,n}(\mathcal{E}_{i},|\mathbf{r}_{1}-\mathbf{r}_{i}|,...,|\mathbf{r}_{N_{\text{c}}}-\mathbf{r}_{i}|)  &= \frac{\rho_{i,n}}{{\sum_{{n'=1}}^{{N_{\text{c}}}}{\rho_{i,n'}}}} ,
\end{split}
\end{align}
with
\begin{align}\label{unnorm_dense}
\begin{split}
{\rho_{i,n}(t)} &= | \psi (\zeta_i,|\mathbf{r}_{n}-\mathbf{r}_{i}|) |^2 = {\frac{\zeta_{i,n}^3}{\pi}e^{-2\zeta_{i,n}(t)|\mathbf{r}_{n}-\mathbf{r}_{i}|}}.
\end{split}
\end{align}
Hence, we  approximate the wavefunction of each bound electron $i$ with a $1s$ hydrogenic wavefunction around each core. In addition, we are distributing the charge of a bound electron $i$ amongst the different cores according to the probability density of electron $i$ with respect to each core. Hence, the total potential that a bound electron $j$ experiences due to a bound electron $i$ is given by
\begin{equation}
\sum_{n=1}^{N_{\text{c}}}C_{i,n}(\mathcal{E}_{i},|\mathbf{r}_{1}-\mathbf{r}_{i}|,...,|\mathbf{r}_{N_{\text{c}}}-\mathbf{r}_{i}|)V_{\text{eff}}(\zeta_{i,n}(t),|\mathbf{r}_{n}-\mathbf{r}_{j}|),
\end{equation}
where $C_{i,n}(\mathcal{E}_{i},|\mathbf{r}_{1}-\mathbf{r}_{i}|,...,|\mathbf{r}_{N_{\text{c}}}-\mathbf{r}_{i}|)$ is a function of the energy of electron $i$ and the distance between electron $i$ and the cores. $C_{i,n}(\mathcal{E}_{i},|\mathbf{r}_{1}-\mathbf{r}_{i}|,...,|\mathbf{r}_{N_{\text{c}}}-\mathbf{r}_{i}|)$ can be more explicitly written as $C_{i,n}(\zeta_{i,1},...,\zeta_{i,N_{\text{c}}},|\mathbf{r}_{1}-\mathbf{r}_{i}|,...,|\mathbf{r}_{N_{\text{c}}}-\mathbf{r}_{i}|)$, but since $\zeta_{i,n} \propto \mathcal{E}_i, $ we simplify the expression by only including $\mathcal{E}_i$.

\subsection{Hamiltonian of the system}\label{Sec::MethodB}
The Hamiltonian of the $N$-body molecular system, comprised of $N_{\text{c}}$ cores and $N$-$N_{\text{c}}$ electrons in the nondipole approximation is given by
{\allowdisplaybreaks
\begin{align}\label{Ham}
&H = \sum_{i=1}^{N}\frac{\left[\mathbf{\tilde{p}}_{i}- Q_i\mathbf{A}(\mathbf{r}_{i},t) \right]^2}{2m_i}+\sum_{n=1}^{N_{\text{c}}}\sum_{j= i +1}^{N}\frac{Q_nQ_j}{|\mathbf{r}_n-\mathbf{r}_j|}  \nonumber \\
&+\sum_{i=N_{\text{c}}+1}^{N-1}\sum_{j=i+1}^{N} \left[ 1-c_{i,j}(t)\right]\frac{Q_i Q_j}{|\mathbf{r}_i-\mathbf{r}_j|} \\
&+\sum_{i=N_{\text{c}}+1}^{N-1}\sum_{j=i+1}^{N}c_{i,j}(t)V_{i,j}\nonumber,
\end{align}}
where ${Q_i}$ is the charge, ${m_i}$ is the mass, $\mathbf{r}_{{i}}$ is the position vector and $\mathbf{\tilde{p}}_{{i}}$ is the canonical momentum vector of particle $i$. The mechanical momentum ${\mathbf{p}_{i}}$ is given by
\begin{equation}\label{eq:mechanical_momentum}
{\mathbf{p}_{i}=\mathbf{\tilde{p}}_{{i}}- {Q_i}\mathbf{A}(\mathbf{r}_{{i}},{t}}).
\end{equation}
The  potential $V_{i,j}$ is given as a sum of effective potentials as follows
{\allowdisplaybreaks
\begin{align}
&V_{i,j} = \sum_{n=1}^{N_{\text{c}}} \Big[ C_{j,n}(\mathcal{E}_{j},|\mathbf{r}_{1}-\mathbf{r}_{j}|,...,|\mathbf{r}_{N_{\text{c}}}-\mathbf{r}_{j}|)\times \nonumber \\
& \times V_{\text{eff}}(\zeta_{j,n}(t),|\mathbf{r}_{n}-\mathbf{r}_{i}|) + C_{i,n}(\mathcal{E}_{i},|\mathbf{r}_{1}-\mathbf{r}_{i}|,...,|\mathbf{r}_{N_{\text{c}}}-\mathbf{r}_{i}|) \times \nonumber\\
& \times V_{\text{eff}}(\zeta_{i,n}(t),|\mathbf{r}_{n}-\mathbf{r}_{j}|) \Big]
\end{align}}
The functions ${c_{i,j}(t)}$ determine whether the full Coulomb interaction or the effective $V_{\text{eff}}(\zeta_{i,n}(t),|\mathbf{r}_{n}-\mathbf{r}_{j}|)$ and $V_{\text{eff}}(\zeta_{j,n}(t),|\mathbf{r}_{n}-\mathbf{r}_{i}|)$ potential interactions are on or off for any pair of electrons $i$ and $j$ during the time propagation. Specifically, the limiting values of ${c_{i,j}(t)}$ are zero and one. The value zero corresponds to the full Coulomb potential being turned on while the effective Coulomb potentials are off. This occurs for a pair of electrons $i$ and $j$ where either electron $i$ or $j$ is quasifree. The value one corresponds to the effective Coulomb potentials $V_{\text{eff}}(\zeta_{i,n}(t),|\mathbf{r}_{n}-\mathbf{r}_{j}|)$ and $V_{\text{eff}}(\zeta_{j,n}(t),|\mathbf{r}_{n}-\mathbf{r}_{i}|)$ being turned on while the full Coulomb potential is off. This occurs when both electrons $i$ and $j$ are bound. For simplicity, we choose ${c_{i,j}(t)}$ to change linearly with time between the limiting values zero and one \cite{Agapi3electron,PhysRevA.107.L041101}. Hence, ${c_{i,j}(t)}$ is defined as follows 
\begin{equation}\label{eqn::zeta_charges_section}
{c_{i,j}(t)} = \left\{
    \begin{array}{ll}
        0 & {c(t)}  \leq 0 \\
        {c(t) } & 0 < {c(t)} < 1 \\
        1 & {c(t)}  \geq 1,
    \end{array}
\right.
\end{equation}
where ${c(t) = \beta (t-t^{i,j}_s)+c_{0},} $ and ${c_0}$ is the value of  ${c_{i,j}(t)} $ just before a switch at time ${t^{i,j}_s}$. A switch at time ${t^{i,j}_s}$ occurs if the interaction between electrons $i, j$ changes from full Coulomb to effective Coulomb potential or vice versa. Every time during propagation that such a switch takes place, we check whether for each pair of electrons the full Coulomb force should be switched on and hence the effective potential switched off or the full Coulomb force should be switched off and the effective potential switched on. The former occurs if at time ${t^{i,j}_s}$ one of the two electrons in a pair of bound electrons changes to being quasifree while the latter occurs if in a pair of a quasifree electron and a bound electron the quasifree electron becomes bound.  At the start of the propagation at time ${t_0},$ ${t^{i,j}_s}$ is equal to ${t_0}$ and ${c_0}$ is one for pairs of electrons that are bound and zero otherwise. To allow  the effective Coulomb potential to be switched on or off in a smooth way, we choose $\beta$ equal to $\pm 0.1;$ plus corresponds to a switch on and minus to a switch off of the effective Coulomb potential.


\subsection{Global regularisation}\label{Regular}
We perform a global regularisation \cite{Heggie1974} to avoid any numerical issues arising from the Coulomb singularities. We previously used this regularisation scheme, for strongly driven H$_2$, to study double  and frustrated double ionization  within the dipole approximation \cite{toolkit2014} as well as nondipole effects in nonsequential double ionization \cite{PhysRevA.103.033115}. Also, we have used this regularisation to study triple ionization in the nondipole approximation in strongly driven Ar \cite{Agapi3electron} and Ne \cite{PhysRevA.107.L041101}. In this scheme, our new coordinates involve the relative position between two particles $i$ and $j$ 
\begin{equation}
\mathbf{q}_{{ij}}=\mathbf{r}_{{i}}-\mathbf{r}_{{j}}
\label{eq:position}
\end{equation}
and their conjugate momenta
\begin{equation}
\boldsymbol{\rho}_{{ij}}=\frac{1}{{N}}\left( \mathbf{\tilde{p}}_{{i}} - \mathbf{\tilde{p}}_{{j}}- \frac{{m_i}-{m_j}}{{M}}\langle \boldsymbol{\rho} \rangle \right) ,
\end{equation}
where
\begin{equation}
\langle \boldsymbol{\rho} \rangle = \sum_{{i=1}}^{{N}}\mathbf{\tilde{p}}_{{i}} \; \; \text{and} \; \; {M}=\sum_{{i=1}}^{{N}}{m_i}.
\end{equation}
The inverse transformation is given by
\begin{equation}\label{position_in_terms_of_q}
\mathbf{r}_{i}=\frac{1}{{M}}\sum_{{j=i+1}}^{{N}}{m_j}\mathbf{{q}}_{{ij}}-\frac{1}{{M}}\sum_{{j=1}}^{{i-1}}{m_j}\mathbf{q}_{{ji}}+ \langle \mathbf{q} \rangle,
\end{equation}
and
\begin{equation}\label{momenta_in_terms_of_rho}
\mathbf{\tilde{p}}_{{i}}=\sum_{{j=i+1}}^{{N}}\boldsymbol{\rho}_{{ij}}-\sum_{{j=1}}^{{i-1}}\boldsymbol{\rho}_{{ji}}+ \frac{{m_i}}{{M}} \langle \boldsymbol{\rho} \rangle,
\end{equation}
where
\begin{equation}
\langle \mathbf{q} \rangle = \frac{1}{{M}} \sum_{{i=1}}^{{N}}{m_i}\mathbf{r}_{{i}}.
\end{equation}

Next, we define a fictitious particle ${k}$ for each pair of particles ${i,j}$ as follows
\begin{equation}
{k(i,j)} = { (i-1)N - \dfrac{i(i+1)}{2}+ j },
\label{eqn::kdef}
\end{equation}
with $j>i$ and the total number of fictitious particles being equal to ${K=N(N-1)/2}$. In addition, we define the parameters $\alpha_{{ik}}$ and $\beta_{{ik}},$ as $\alpha_{{ik}}=1,\beta_{{ik}}={m_j/M}$ and $\alpha_{{jk}}=-1,\beta_{{jk}}=-{m_i/M}$ when ${k=k(i,j)}$, otherwise $\alpha_{{ik}}=\beta_{{ik}}=0$. Given the above, Eqs. (\ref{position_in_terms_of_q})  and (\ref{momenta_in_terms_of_rho}) take the following simplified form
\begin{equation}\label{momenta_in_terms_of_rho_v2}
\mathbf{\tilde{p}}_{{i}}=\sum_{{k=1}}^{K}\alpha_{ik}\boldsymbol{\rho}_{{k}}+ \frac{{m_i}}{{M}} \langle \boldsymbol{\rho} \rangle,
\end{equation}
and
\begin{equation}\label{positions_in_terms_of_rho_v2}
\mathbf{r}_{{i}}=\sum_{{k=1}}^{K}\beta_{ik}\mathbf{q}_{{k}}+ \langle \mathbf{q} \rangle.
\end{equation}

\subsection{Derivation of the time derivative of the effective charges}\label{sec:derivation_charges_eff}
The Hamiltonian in \eq{Ham} depends not only on positions,
momenta, and time but also on the effective charges. Moreover,
the Hamiltonian depends on time through the vector
potential as well as through the effective charges that are
time dependent. Since the effective charge of electron $j$ is proportional
to the energy $\mathcal{E}_j(t)$ [see \eq{eq:energy_of_electron_j}], it follows that we must obtain the derivative with respect to time of $\mathcal{E}_j(t)$. This is necessary at any time during propagation if at least two electrons are bound. Following the same procedure as in Ref. \cite{Agapi3electron}, we calculate the time derivative of the energy of electron $j$. To do so, we apply the chain rule in \eq{eq:energy_of_electron_j} and obtain 
{\allowdisplaybreaks
\begin{widetext}
\begin{align}\label{eq:total_time_derivative}
\begin{split}
\mathcal{\dot{E}}_{j}(t) &=\frac{\partial \mathcal{E}_{j}(t)}{\partial \mathbf{r}_{{j}}}\cdot\dot{\mathbf{r}}_{j} + \frac{\partial \mathcal{E}_{j}(t)}{\partial \mathbf{\tilde{p}}_{j}}\cdot \dot{\tilde{\mathbf{p}}}_{j} + \sum_{n=1}^{N_{\text{c}}}\frac{\partial \mathcal{E}_{j}(t)}{\partial \mathbf{r}_{n}}\cdot\dot{\mathbf{r}}_{n} +\sum_{\substack{\;{i=N_{\text{c}}+1} \\ i \neq j}}^{{N}}{\frac{\partial \mathcal{E}_{j}(t)}{\partial \mathbf{r}_{i}}\cdot\dot{\mathbf{r}}_{i}}+ \sum_{\substack{\;{i=N_{\text{c}}+1} \\ {i} \neq {j}}}^{{N}}\frac{\partial \mathcal{E}_{j}(t)}{\partial \mathcal{E}_i} \dot{\mathcal{E}_i}  + {\frac{\partial \mathcal{E}_{j}(t)}{\partial t}}   \\
&=  \frac{\partial \left[\mathcal{E}_{j}(t) - H\right]}{\partial \mathbf{r}_{{j}}}\cdot\dot{\mathbf{r}}_{{j}} + \sum_{n=1}^{N_{\text{c}}}\frac{\partial \mathcal{E}_{j}(t)}{\partial \mathbf{r}_{n}}\cdot\dot{\mathbf{r}}_{n }+ \sum_{\substack{\;{i={N_{\text{c}}}+1} \\ {i} \neq {j}}}^{{N}}{\frac{\partial \mathcal{E}_{j}(t)}{\partial \mathbf{r}_{{i}}}\cdot\dot{\mathbf{r}}_{{i} }} +{\sum_{\substack{\;{i={N_{\text{c}}}+1} \\ {i} \neq {j}}}^{{N}}}\frac{\partial \mathcal{E}_{j}(t)}{\partial \mathcal{E}_i} \dot{\mathcal{E}_i}  + {\frac{\partial \mathcal{E}_{j}(t)}{\partial t}},
\end{split}
\end{align}
\end{widetext}
}
\noindent where we use ${\dot{\mathbf{r}}_j = \frac{\partial \mathcal{E}_{j}(t)}{\partial \mathbf{\tilde{p}}_{{j}}}}$ and $\dot{\tilde{\mathbf{p}}}_{{j}}=-{\frac{\partial H }{\partial \mathbf{r}_j }} $.  In Appendix \ref{Appendix_A}, we derive each of the  terms in the chain rule in Eq. \eqref{eq:total_time_derivative}.
Furthermore, we group together all the terms in Eq. \eqref{eq:total_time_derivative} that do not depend on $\dot{\mathcal{E}}_{i}$ as follows
\begin{align}\label{eq:linear_equations}
\begin{split}
\mathcal{\dot{E}}_{j}(t) &= { f_j + {\sum_{\substack{\;{i=N_c+1} \\ {i} \neq {j}}}^{{N}} g_{j,i} \dot{\mathcal{E}_i}  }},
\end{split}
\end{align}
where 
{\allowdisplaybreaks
\begin{widetext}
\begin{align}\label{eq:g_ji}
\begin{split}
g_{j,i}&=  \sum_{n=1}^{N_{\text{c}}}c_{i,j}(t)\left[C_{i,n}(\mathcal{E}_{i},|\mathbf{r}_{1}-\mathbf{r}_{i}|,...,|\mathbf{r}_{N_{\text{c}}}-\mathbf{r}_{i}|) \frac{\partial V_{\text{eff}}(\zeta_{i,n},|\mathbf{r}_{n}-\mathbf{r}_{j}|)}{\partial \zeta_{i,n}}  \frac{\partial \zeta_{i,n}}{\partial \mathcal{E}_{i}}\right. \\
&\left.+V_{\text{eff}}(\zeta_{i,n},|\mathbf{r}_{n}-\mathbf{r}_{j}|)\left(\sum_{b=1}^{{{N_{\text{c}}}}} \frac{\partial C_{i,n}(\mathcal{E}_{i},|\mathbf{r}_{1}-\mathbf{r}_{i}|,...,|\mathbf{r}_{N_{\text{c}}}-\mathbf{r}_{i}|)}{\partial \zeta_{i,b}} \frac{\partial \zeta_{i,b}}{\partial \mathcal{E}_{i}}\right)\right],
\end{split}
\end{align}
\end{widetext}
}
and
\begin{equation}\label{eqn::zeta_and_energy_derivative}
\frac{\partial \zeta_{i,n}}{\partial \mathcal{E}_{i}} = \left\{
    \begin{array}{ll}
        0 & \mathcal{E}_{i}(t)  \leq \mathcal{E}^{n}_{\mathrm{1 s}}\\
        \left({Q_n} / \mathcal{E}^{n}_{1 s}\right)  & \mathcal{E}^{n}_{\mathrm{1 s}} < \mathcal{E}_{i}(t)  < 0 \\
        0 & \mathcal{E}_{i}(t)  \geq 0.
    \end{array}
\right.
\end{equation}
 The derivatives  $\frac{\partial C_{i,n}(\mathcal{E}_{i},|\mathbf{r}_{1}-\mathbf{r}_{i}|,...,|\mathbf{r}_{N_{\text{c}}}-\mathbf{r}_{i}|)}{\partial \zeta_{i,b}}$ and $\frac{\partial V_{\text{eff}}(\zeta_{i,n},|\mathbf{r}_{n}-\mathbf{r}_{j}|)}{\partial \zeta_{i,n}}$ are  obtained in \eq{Cim_constant_derivative_zeta} and \eq{eq:Veff_derivative_zeta} of Appendix \ref{Appendix_B}, respectively. These derivatives are  obtained in terms of the relative coordinates $\mathbf{q}_{k}$, since we propagate in the regularised coordinates system. We also find that 
{\allowdisplaybreaks
\begin{widetext}
\begin{align}\label{eq:f_j}
\begin{split}
{f_{j}}&= {\sum_{{i}={N_{\text{c}}}+1}^{{N-1}}\sum_{m={i}+1}^{{N}}  [1-c_{i,m}(t)] {\frac{Q_i Q_m(\mathbf{r}_{i}-\mathbf{r}_{m})}{|\mathbf{r}_{i}-\mathbf{r}_{m}|^3}\left(\delta_{i,j} - \delta_{m,j}\right)  \cdot \dot{\mathbf{r}}_{{j}}  }   }\\
&- {\sum_{\substack{\;{i={N_{\text{c}}}+1} \\ {i} \neq {j}}}^{{N}}}{\sum_{n=1}^{N_{\text{c}}}  c_{i,j}(t) \frac{\partial C_{j,n}(\mathcal{E}_{j},|\mathbf{r}_{1}-\mathbf{r}_{j}|,...,|\mathbf{r}_{N_{\text{c}}}-\mathbf{r}_{j}|)}{\partial \mathbf{r}_j}V_{\text{eff}}(\zeta_{j,n},|\mathbf{r}_{n}-\mathbf{r}_{i}|)\cdot \mathbf{\dot{r}}_{j} }\\
&+ \sum_{\substack{\;{i={N_{\text{c}}}+1} \\ {i} \neq {j}}}^{N} \sum_{n=1}^{N_{\text{c}}}c_{i,j}(t)\left[ C_{i,n}(\mathcal{E}_{i},|\mathbf{r}_{1}-\mathbf{r}_{i}|,...,|\mathbf{r}_{N_{\text{c}}}-\mathbf{r}_{i}|)\frac{\partial V_{\text{eff}}(\zeta_{i,n},|\mathbf{r}_{n}-\mathbf{r}_{j}|)}{\partial \mathbf{r}_n} \right. \\
&\left. +\sum_{b=1}^{N_{\text{c}}} V_{\text{eff}}(\zeta_{i,b},|\mathbf{r}_{b}-\mathbf{r}_{j}|)\frac{\partial C_{i,b}(\mathcal{E}_{i},|\mathbf{r}_{1}-\mathbf{r}_{i}|,...,|\mathbf{r}_{N_{\text{c}}}-\mathbf{r}_{i}|)}{\partial \mathbf{r}_n}\right]  \cdot \dot{\mathbf{r}}_{n}   \\
&+ {\sum_{n=1}^{N_{\text{c}}}\left[-\frac{Q_n Q_j(\mathbf{r}_{n}-\mathbf{r}_{j})}{|\mathbf{r}_{n}-\mathbf{r}_{j}|^3} \right]  \cdot \dot{\mathbf{r}}_{n}   }+\sum_{\substack{\;{i={N_{\text{c}}}+1} \\ {i} \neq {j}}}^{{N}} \sum_{n=1}^{ N_{\text{c}}} c_{i,j}(t)\left[V_{\text{eff}}(\zeta_{i,n},|\mathbf{r}_{n}-\mathbf{r}_{j}|)\frac{\partial C_{i,n}(\mathcal{E}_{i},|\mathbf{r}_{1}-\mathbf{r}_{i}|,...,|\mathbf{r}_{N_{\text{c}}}-\mathbf{r}_{i}|)}{\partial \mathbf{r}_i}\right]  \cdot \dot{\mathbf{r}}_{{i}}  \\
&+ {\sum_{\substack{\;{i={N_{\text{c}}}+1} \\ {i} \neq {j}}}^{{N}}}{\sum_{n=1}^{N_{\text{c}}} \dot{c}_{i,j}(t) C_{i,n}(\mathcal{E}_{i},|\mathbf{r}_{1}-\mathbf{r}_{i}|,...,|\mathbf{r}_{N_{\text{c}}}-\mathbf{r}_{i}|)V_{\text{eff}}(\zeta_{i,n},|\mathbf{r}_{n}-\mathbf{r}_{j}|) }-{{Q_j}\mathbf{r}_j \cdot \dot{\mathbf{E}}(\mathbf{r}_j,{t})},
\end{split}
\end{align}
\end{widetext}}
where $\dot{\mathbf{r}}_{n}$ = $\frac{\mathbf{p}_n}{m_n}$, $\dot{\mathbf{r}}_{i}$=$\frac{\mathbf{p}_i}{m_i}$ and $\dot{\mathbf{r}}_{j}$=$\frac{\mathbf{p}_j}{m_j}$.The derivatives of $ C_{j,n}$ and $V_{\text{eff}}$ in Eq. \eqref{eq:f_j}   can be found in Eqs. \eqref{Cim_constant_derivative_ri}, \eqref{Cim_constant_derivative_rn} and \eqref{eq:Veff_derivative_rn} in Appendix \ref{Appendix_B}. These derivatives and all the terms in Eqs. \eqref{eq:g_ji} and \eqref{eq:f_j} are obtained in terms of the relative coordinates $\mathbf{q}_{k},$ since we propagate in the regularised coordinates system. For each electron we obtain an equation similar to \eq{eq:linear_equations}. Hence, at any time during time propagation, we solve a system of $N$-$N_{c}$ equations to obtain the derivative in terms of the energy of each electron, so that it does not depend on the derivatives of the other electron energies. We solve this system of linear equations using  Cramers rule \cite{cramer1750introduction,muir1960theory}.

\subsection{Hamilton's equations of motion}
Substituting Eqs. \eqref{eq:position} and \eqref{momenta_in_terms_of_rho_v2} in \eq{Ham}, we find that the Hamiltonian in regularized coordinates is given by
\begin{align}\label{Hamiltonian_in_transformed_coordinates}
 \begin{split}
H =& {\sum_{{k,k'=1}}^{{K}}{T}_{{kk'}}\boldsymbol{\rho}_{{k}}\boldsymbol{\rho}_{{k'}}+\frac{\langle \boldsymbol{\rho}\rangle^2}{2M}} + \sum_{{k}={1}}^{{K}}[1-{c_{k}(t)}]\frac{{U_k}}{{q_k}}\\
&{+\sum_{{i=1}}^{{N}}\frac{{Q^2_i}}{2{m_i}}\mathbf{A}^2\left( \mathbf{r}_{{i}} ,{t} \right)  - \sum_{{i=1}}^{{N}}\frac{{Q_i}}{{m_i}}\mathbf{\tilde{p}}_{{i}} \cdot \mathbf{A}\left( \mathbf{r}_{{i}} ,{t} \right)}\\
&+\sum_{{k=1}}^{{K}}\sum_{{n=1}}^{{N_{\text{c}}}}{c_{k}(t)}{V_{k,n}},
\end{split}
\end{align}
where $U_{k(i,j)}=Q_iQ_j.$ The term ${V_{k(i,j),n}}$ is now given by 
\begin{align}\label{veff_molecular}
\begin{split}
&V_{k(i,j),n} = \\
&C_{j,n}(\mathcal{E}_{j},|\mathbf{r}_{1}-\mathbf{r}_{j}|,...,|\mathbf{r}_{N_{\text{c}}}-\mathbf{r}_{j}|)V_{\text{eff}}(\zeta_{j,n}(t),|\mathbf{r}_{n}-\mathbf{r}_{i}|) \\
&+ C_{i,n}(\mathcal{E}_{i},|\mathbf{r}_{1}-\mathbf{r}_{i}|,...,|\mathbf{r}_{N_{\text{c}}}-\mathbf{r}_{i}|)V_{\text{eff}}(\zeta_{i,n}(t),|\mathbf{r}_{n}-\mathbf{r}_{j}|) 
\end{split}
\end{align}
and $\mathbf{\tilde{p}}$, $\mathbf{r},$ are expressed in terms of $\boldsymbol{\rho}$ and $\mathbf{q}$ via  Eqs. \eqref{momenta_in_terms_of_rho_v2} and \eqref{positions_in_terms_of_rho_v2}. Moreover, we set  ${c_k(t)=0}$ when $k$ corresponds to the relative distance between an electron and a  core. This is the case, since in our model the Coulomb potential between an electron and a core is given by the full Coulomb potential. Using \eq{Hamiltonian_in_transformed_coordinates}, we find that Hamilton's equations of motion are given by 
\begin{align}\label{eq:new_Equations_of_motion_eff}
\begin{split}
\frac{{d}\mathbf{q}_{{k}}}{{dt}}&=2\sum_{{k'}=1}^{{K}}{T}_{{kk'}}\boldsymbol{\rho}_{{k'}}-\sum_{{i=1}}^{{N}}\frac{{Q_i}}{{m_i}} \alpha_{{ik}}\mathbf{A}\left( \mathbf{r}_{{i}} ,{t} \right),   \\
\frac{{d}\langle \mathbf{q} \rangle}{{dt}}&=\dfrac{1}{{M}}\langle \boldsymbol{\rho} \rangle -\sum_{{i=1}}^{{N}}\frac{{Q_i}}{{M}} \mathbf{A}\left( \mathbf{r}_{{i}} ,{t} \right),  \\
\frac{{d}\boldsymbol{\rho}_{{k}}}{{dt}}&= {[1-c_k(t)] }\frac{{U_k}\mathbf{q}_{{k}}}{{q^3_k}} -{ \sum_{k'=1}^{K}  c_{k'}(t)   h_{k}^{k'}}\\
&+ \sum_{{i=1}}^{{N}}\frac{{Q_i}}{{m_i}} \left[ \mathbf{\tilde{p}}_{{i}} - {Q_i}\mathbf{A}\left( {\mathbf{r}}_{{i}},{t} \right)\right]\cdot\dfrac{\partial\mathbf{A}\left( {\mathbf{r}}_{{i}},{t} \right)}{\partial \mathbf{q_k}},  \\
\frac{{d}\langle \boldsymbol{\rho}\rangle}{{dt}}&=\sum_{{i=1}}^{{N}}\frac{{Q_i}}{{m_i}} \left[ \mathbf{\tilde{p}}_{{i}} - {Q_i}\mathbf{A}\left( {\mathbf{r}}_{{i}},{t} \right)\right]\cdot\dfrac{\partial\mathbf{A}\left( {\mathbf{r}}_{{i}},{t} \right)}{\partial \langle \mathbf{q} \rangle},
\end{split}
\end{align}
where
\begin{align}\label{eq:general_form_of_hk}
\begin{split}
{ \sum_{k'=1}^{K}  c_{k'}(t)   h_{k}^{k'}} &=  {\sum_{k'=1}^{K}\sum_{{n=1}}^{N_{\text{c}}}c_{k'}(t)\dfrac{\partial V_{k',n}}{\partial \mathbf{q}_{k}}}.
\end{split}
 \end{align} 
 We find $\dfrac{\partial V_{k',n}}{\partial \mathbf{q}_{k}}$  to be given by
 \begin{widetext}
 \begin{align}\label{eq:new_Equations_of_motion_eff_more}
\begin{split}
&\frac{\partial V_{k'(i',j'),n}}{\partial \mathbf{q}_{k(i,j)}}\\
&= \delta_{i',j} \left[ \delta_{n,i} C_{j',n}(\mathcal{E}_{j'},|\mathbf{r}_{1}-\mathbf{r}_{j'}|,...,|\mathbf{r}_{N_{\text{c}}}-\mathbf{r}_{j'}|)\dfrac{\partial V_{\text{eff}}(\zeta_{j',n},| \mathbf{r}_n - \mathbf{r}_{i'} |)}{\partial \mathbf{q_k}} + \dfrac{\partial C_{i',n}(\mathcal{E}_{i'},|\mathbf{r}_{1}-\mathbf{r}_{i'}|,...,|\mathbf{r}_{N_{\text{c}}}-\mathbf{r}_{i'}|)}{\partial \mathbf{q_k}}V_{\text{eff}}(\zeta_{i',n},| \mathbf{r}_n - \mathbf{r}_{j'} |) \right] \\
&+\delta_{j',j} \left[ \delta_{n,i} C_{i',n}(\mathcal{E}_{i'},|\mathbf{r}_{1}-\mathbf{r}_{i'}|,...,|\mathbf{r}_{N_{\text{c}}}-\mathbf{r}_{i'}|)\dfrac{\partial V_{\text{eff}}(\zeta_{i',n},| \mathbf{r}_n - \mathbf{r}_{j'} |)}{\partial \mathbf{q_k}} + \dfrac{\partial C_{j',n}(\mathcal{E}_{j'},|\mathbf{r}_{1}-\mathbf{r}_{j'}|,...,|\mathbf{r}_{N_{\text{c}}}-\mathbf{r}_{j'}|)}{\partial \mathbf{q_k}}V_{\text{eff}}(\zeta_{j',n},| \mathbf{r}_n - \mathbf{r}_{i'} |) \right].
\end{split}
 \end{align}
 \end{widetext}
Thus, we obtain for ${ h_{k}^{k'}}$
 \begin{widetext}
\begin{align}\label{eq:derivation_of_h_m}
\begin{split}
&h_{k}^{k'} = \sum_{n=1}^{N_{\text{c}}}{\frac{\partial V_{k'(i',j'),n}}{\partial \mathbf{q}_{k(i,j)}}} \\
&= { \delta_{i',j} \left[C_{j',i}(\mathcal{E}_{j'},|\mathbf{r}_{1}-\mathbf{r}_{j'}|,...,|\mathbf{r}_{N_{\text{c}}}-\mathbf{r}_{j'}|)\dfrac{\partial V_{\text{eff}}(\zeta_{j',i},| \mathbf{r}_{i} - \mathbf{r}_{i'} |)}{\partial \mathbf{q_k}} + \sum_{{n=1}}^{N_\text{c}}\dfrac{\partial C_{i',n}(\mathcal{E}_{i'},|\mathbf{r}_{1}-\mathbf{r}_{i'}|,...,|\mathbf{r}_{N_{\text{c}}}-\mathbf{r}_{i'}|)}{\partial \mathbf{q_k}}V_{\text{eff}}(\zeta_{i',n},| \mathbf{r}_{n} - \mathbf{r}_{j'} |) \right]} \\
&+{ \delta_{j',j} \left[C_{i',i}(\mathcal{E}_{i'},|\mathbf{r}_{1}-\mathbf{r}_{i'}|,...,|\mathbf{r}_{N_{\text{c}}}-\mathbf{r}_{i'}|)\dfrac{\partial V_{\text{eff}}(\zeta_{i',i},| \mathbf{r}_{i} - \mathbf{r}_{j'} |)}{\partial \mathbf{q_k}} + \sum_{{n=1}}^{N_{\text{c}}}\dfrac{\partial C_{j',n}(\mathcal{E}_{j'},|\mathbf{r}_{1}-\mathbf{r}_{j'}|,...,|\mathbf{r}_{N_{\text{c}}}-\mathbf{r}_{j'}|)}{\partial \mathbf{q_k}}V_{\text{eff}}(\zeta_{j',n},| \mathbf{r}_{n} - \mathbf{r}_{i'} |) \right]},
\end{split}
 \end{align}
  \end{widetext}
where the derivatives of $V_{\text{eff}}$ and $C$ can be found in Appendix \ref{Appendix_B} in  \eq{eq:Veff_derivative_qk} and \eq{Cim_constant_derivative_qk}, respectively. Note that when $N_{\text{c}}=1$,  Hamilton's equations in \eq{eq:new_Equations_of_motion_eff} reduce to the corresponding equations for an atom \citep{Agapi3electron}.

\subsection{Time derivative of the Hamiltonian}
During propagation, besides other criteria used that we do not detail here, we check the accuracy of our propagation also by comparing the propagated Hamiltonian with the Hamiltonian obtained by substituting the propagated variables. The time derivative of the Hamiltonian in \eq{Hamiltonian_in_transformed_coordinates} is given by
\begin{align}
\begin{split}
&\frac{dH}{dt} = \sum_{k=1}^{K} \left(\frac{\partial H}{\partial \mathbf{q}_k} \dot{\mathbf{q}}_k + \frac{\partial H}{\partial \boldsymbol{\rho}_k} \dot{\boldsymbol{\rho}}_k  \right) + \frac{\partial H}{\partial \langle \mathbf{q} \rangle} \langle\dot{\mathbf{q}} \rangle+ \frac{\partial H}{\partial \langle\boldsymbol{\rho} \rangle} \langle\dot{\boldsymbol{\rho}}\rangle \\& + \frac{\partial H}{ \partial t}
\\&= \sum_{k=1}^{K} \left(-\dot{\boldsymbol{\rho}}_k \dot{\mathbf{q}}_k + \dot{\mathbf{q}}_k \dot{\boldsymbol{\rho}}_k  \right) - \langle\dot{\boldsymbol{\rho}}\rangle \langle\dot{\mathbf{q}} \rangle+ \langle\dot{\mathbf{q}} \rangle \langle\dot{\boldsymbol{\rho}}\rangle  + \frac{\partial H}{ \partial t}\\
&= \frac{\partial H}{ \partial t}
\end{split}
\end{align}
The partial derivative of the Hamiltonian with respect to time is given as follows
\begin{align}
\begin{split}
& \frac{\partial H}{ \partial t} = \sum_{i=N_{\text{c}+1}}^{N-1} \sum_{j=i+1}^{{N}}g_{j,i}\dot{\mathcal{E}}_i +\sum_{{k=1}}^{{K}}\sum_{n=1}^{N_{\text{c}}}{\dot{c}_{k}(t)}{V_{k,n}} \\
&-\sum_{{k} ={1}}^{{K}}{\dot{c}_{k}(t)}\frac{{U_k}}{{q_k}}+  \sum_{{i=1}}^{{N}}\frac{{Q_i}^2}{{m_i}}\mathbf{A}\left( \mathbf{r}_{{i}} ,{t} \right)\frac{\partial \mathbf{A}\left( \mathbf{r}_{{i}} ,{t} \right)}{\partial t} \\
&-\sum_{{i=1}}^{{N}}\frac{{Q_i}}{{m_i}}\tilde{\mathbf{p}}_{{i}} \cdot\frac{\partial \mathbf{A}\left( \mathbf{r}_{{i}} ,{t} \right)}{\partial t}      \\
&= \sum_{i=N_{\text{c}+1}}^{N-1} \sum_{j=i+1}^{{N}}g_{j,i}\dot{\mathcal{E}}_i +\sum_{{k=1}}^{{K}}\sum_{n=1}^{N_{\text{c}}}{\dot{c}_{k}(t)}{V_{k,n}} \\
&-\sum_{{k} ={1}}^{{K}}{\dot{c}_{k}(t)}\frac{{U_k}}{{q_k}} +{ \sum_{{i=1}}^{{N}}\frac{{Q_i}}{{m_i}}\mathbf{p}_{{i}} \cdot \mathbf{E}\left( \mathbf{r}_{{i}} ,{t} \right)}
\end{split}
\end{align}

\subsection{Tunneling during propagation}
During time propagation, we allow for each bound electron to tunnel at the classical turning points along the axis of the electric field using the Wentzel-Kramers-Brillouin (WKB) approximation \cite{WKB}. For the transmission probability we use the WKB formula for transmission through a potential barrier \cite{WKB},
\begin{equation}\label{EQ:Notrootz}
\mathrm{T \approx \exp \left(-2 \int_{r_a}^{r_b}\left[2\left(V_{\mathrm{tun}}\left(r, t_{\mathrm{tun}}\right)-\epsilon_i\right)\right]^{1 / 2} d r\right)}
\end{equation}
with $V_{\mathrm{tun}}\left(r, t_{\mathrm{tun}}\right)$ the potential a bound electron $i$ can tunnel through given by
\begin{align}
\begin{split}
&V_{\mathrm{tun}}\left(r, t_{\mathrm{tun}}\right) = \sum_{n=1}^{N_{\text{c}}} \dfrac{Q_n Q_i}{ | \mathbf{r}_{n} - \mathbf{r}_{i} |} - Q_i\mathbf{r}_{i} \cdot \mathbf{E}\left(\mathbf{r}_{i}, t_{\mathrm{tun}}\right) \\
&+\sum_{n=1}^{N_{\text{c}}}\sum_{\substack{\;{j=N_{\text{c}}+1} \\ {j} \neq {i}}}^{N} c_{i,j}(t_{\mathrm{tun}})C_{j,n}(\mathcal{E}_{j},|\mathbf{r}_{1}-\mathbf{r}_{j}|,...,|\mathbf{r}_{N_{\text{c}}}-\mathbf{r}_{j}|) \times \\
&\times V_{\text{eff}}(\zeta_{j,n}(t_{\mathrm{tun}}),|\mathbf{r}_{n}-\mathbf{r}_{i} |).
\end{split}
\end{align}
$\epsilon_i$ is the energy of the electron at the time of tunneling, $t_{\text{tun}}$, and $r_a$ and $r_b$ are the classical turning points. We find that accounting for tunneling during time propagation is necessary in order to accurately describe phenomena related to enhanced ionization \cite{Enhanced1,Enhanced2,Enhanced3,Enhanced4,Enhanced5} during the fragmentation of strongly driven molecules.

\subsection{Definition of quasifree and bound electron}
In the ECBB model, the interaction between a pair of electrons where at least one is quasifree is described with the full Coulomb potential. Effective Coulomb potentials are used to describe the interaction between bound electrons. At the start of  time propagation, the tunneling electron  is considered quasifree and the other two electrons  are  bound. We  decide on the fly, during time propagation, whether an electron is quasifree or bound using  a simple set of criteria described briefly below \cite{Agapi3electron}.  

A quasifree electron $i$  transitions to bound if, following a minimum approach to the cores, the position of the electron along the field axis is influenced more by the cores than the electric field. We assume that the electron is influenced more by the cores if its position along the electric field has at least two extrema of the same kind in a time interval less than half a period of the laser field.  The minimum approach to the cores is identified by a maximum in the Coulomb potential of the quasifree electron with the cores. Also, at the end of the laser pulse, if the   quasifree electron has negative compensated energy \cite{Leopold_1979}, this electron transitions to  bound. In our studies, we use a compensated energy of an electron $i$ that includes the effective potentials as well and  
is given by
\begin{align}\label{eqn::compensated1}
\begin{split}
&\varepsilon^{\text{comp}}_{i}(t)=  \frac{\mathbf{\tilde{p}}^2_{i} }{2m_i} + \sum_{n=1}^{N_{\text{c}}}\frac{Q_n Q_i}{|\mathbf{r}_1-\mathbf{r}_i|}  +\sum_{n=1}^{N_{\text{c}}}\sum_{\substack{\;{j=N_{\text{c}}+1} \\ {j} \neq {i}}}^{N} c_{i,j}(t) \times\\
&\times  C_{j,n}(\mathcal{E}_{j},|\mathbf{r}_{1}-\mathbf{r}_{j}|,...,|\mathbf{r}_{N_{\text{c}}}-\mathbf{r}_{j}|)  V_{\text{eff}}(\zeta_{j,n}(t),|\mathbf{r}_{n}-\mathbf{r}_{i} |).
\end{split}
\end{align}
 A bound electron $i$ transitions to quasifree if either of the following two conditions is satisfied: (i) the compensated energy of the bound electron converges to a positive value or (ii) the magnitude of the total Coulomb potential of the electron $i$ with the cores is smaller than a threshold value and it  continuously decreases. The criteria are discussed in detail and illustrated in  Ref. \cite{Agapi3electron}.


\subsection{Initial conditions}
\subsubsection{Nuclei}
\begin{figure}[b]
\centering
\includegraphics[width=\linewidth]{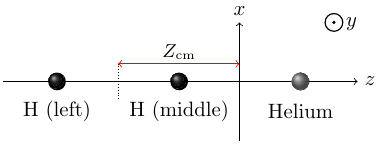}
\caption{Schematic illustration of the molecule under consideration, $\mathrm{HeH_{2}^{+}},$ at the time $t_0$ when we initialize our system. The origin of the coordinate system is set to be the center of mass of the molecule. $\mathrm{Z_{\text{cm}}}$ is the distance between the center of mass of the molecule and the middle of the distance between H left and H middle.}
\label{Fig:Molecule}
\end{figure}

In the initial state of $\mathrm{HeH_{2}^{+}}$, all three atoms are placed along the $z$ axis. The two hydrogen atoms are at -3.09 a.u. and -1.02 a.u., respectively, and the helium atom is at 1.04 a.u., with the origin of the coordinate system set to be the center of mass of the molecule. We refer to H farther away from He as H left and the one closest to He as H middle, see \fig{Fig:Molecule}. We compute the distance between the two hydrogen atoms and the hydrogen and helium atoms using the quantum chemistry package MOLPRO \cite{MOLPRO_brief}, employing the Hartree-Fock method with the augcc-pV5Z basis set. The Hartree-Fock method overestimates by a small amount the distance between the hydrogen and the helium atoms \cite{Palmieri2000}. However, we employ this method for consistency with the Hartree-Fock wave functions that we use in the potential energy terms involved in computing the exit point of the tunnel-ionizing electron \cite{toolkit2014}. All three nuclei are initiated at rest.

\subsubsection{Tunnel-ionizing electron}
The electric field is along the axis of the linear molecule, i.e., the $z$ axis with a field strength within the below-the-barrier ionization regime. As a result, one electron (electron 1) tunnel ionizes at time $t_0$ through the field-lowered Coulomb potential. We employ a quantum-mechanical calculation to compute this ionization rate as described in Ref. \cite{PhysRevA.103.043109}. 
We find $t_0$, using importance sampling \cite{ROTA1986123} in the time interval $[-2\tau,2\tau]$ where the electric field is nonzero; $\tau$ is the full width at half maximum of the pulse duration in intensity. The importance sampling distribution is given by the ionization rate. We assume that electron 1 exits along the direction of the laser field; for details on the exit point, see Ref. \cite{toolkit2014}. We compute the first ionization energy of $\mathrm{HeH_{2}^{+}},$ with MOLPRO and find it equal to 1.02 a.u. The tunnel ionizing electron exits the field-lowered Coulomb barrier with a zero momentum along the direction of the field. The transverse electron momentum is given by a Gaussian distribution. The latter arises from standard tunneling theory \cite{Delone:91,Delone_1998,PhysRevLett.112.213001} and represents the Gaussian-shaped filter with an intensity-dependent width.

\subsubsection{Microcanonical distribution}
\begin{figure}[b]
\centering
\includegraphics[scale=0.3]{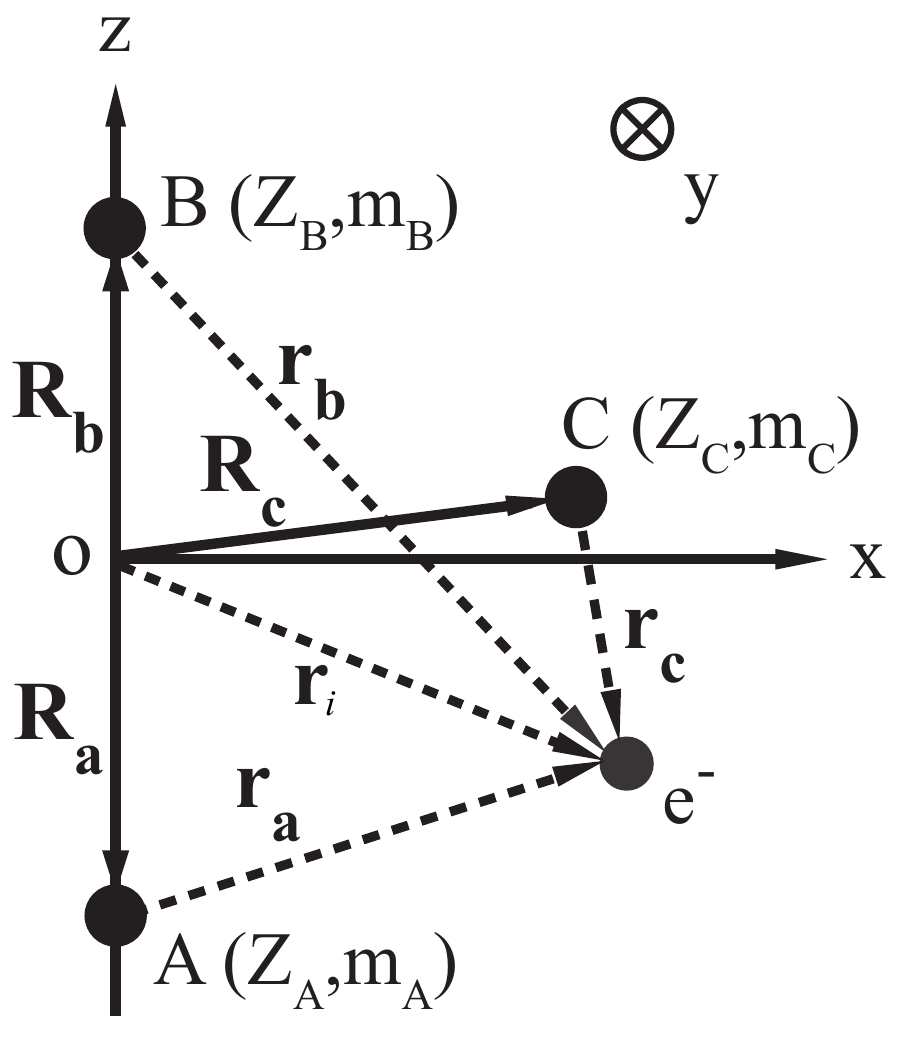}
\caption{The configuration of the triatomic molecule we use to set-up the microcanonical distribution. The origin of the coordinate system is set to be the middle of the distance between the A and B nuclei.}
\label{Fig:Molecule_microcanonical}
\end{figure}

In the ECBB model we obtain the initial position and momentum of each bound electron $i$ at time $t_0$ using a microcanonical distribution with an energy 
\begin{equation}\label{eq:potential_energy_of_electron_ioi}
\mathcal{E}_i(t_0) = \frac{\mathbf{{p}}_{i}^2}{2m_i} + W_i,
\end{equation}
where 
\begin{align}\label{eq:potential_energy_of_electron_i}
\begin{split}
W_i &= \sum_{n=1}^{N_{\text{c}}}\frac{Q_n Q_i}{|\mathbf{r}_{n}-\mathbf{r}_{i}|}   \\
&+ \sum_{n=1}^{N_{\text{c}}}\sum_{\substack{\;{j=N_{\text{c}}+1} \\ {i} \neq {j}}}^{N} c_{i,j}(t)C_{j,n}(\mathcal{E}_j,|\mathbf{r}_{1}-\mathbf{r}_{j}|,...,|\mathbf{r}_{N_{\text{c}}}-\mathbf{r}_{j}|) \times\\
&\times V_{\text{eff}}(\zeta_{j,n}(t),|\mathbf{r}_{n}-\mathbf{r}_{i}|).
\end{split}
\end{align}
We take the energy of each electron to be equal to $-I_{p,2},$ with $I_{p,2},$ being the second ionization potential of the molecule under consideration. We note that the potential energy $W_i$ of each electron $i$ in Eqs. \eqref{eq:potential_energy_of_electron_ioi} and \eqref{eq:potential_energy_of_electron_i} depends not only on the coordinates of the bound electron $i$ but also on the coordinates of all other bound electrons. This dependence is due to the function $C_{j,n}(\mathcal{E}_j,|\mathbf{r}_{1}-\mathbf{r}_{j}|,...,|\mathbf{r}_{N_{\text{c}}}-\mathbf{r}_{j}|).$ Hence, the microcanonical distributions of all bound electrons are interrelated, unlike the case for atoms \cite{Agapi3electron}. For a triatomic molecule, as is $\mathrm{HeH_{2}^{+}}$ (the molecule considered here), indices 1-3 are reserved for the nuclei, index 4 for the electron that tunnels in the initial state and indices 5 and 6 for the bound electrons. The microcanonical distribution for the two bound electrons is given by
\begin{align}\label{eq:multie_micro}
\begin{split}
&f(\mathbf{r}_5, \mathbf{p}_5, \mathbf{r}_6, \mathbf{p}_6)=\\
&=\mathcal{N} \prod_{i=5}^{6} \delta\left[\frac{p_i^2}{2}+W_i(\lambda_5, \mu_5, \phi_5,\lambda_6, \mu_6, \phi_6) \ - (-I_{p,2})\right],
\end{split}
\end{align}
where $N$ is a normalisation constant. To set-up the microcanonical distribution we place the origin of our coordinate system in the middle of the distance between the nuclei A and B, see \fig{Fig:Molecule_microcanonical}. As discussed at the end of this subsection, once we obtain the initial conditions for the position of the electrons in this coordinate system, we shift the positions with respect to the center of mass of the triatomic molecule. Moreover, $\lambda,\mu$ are the confocal elliptical coordinates defined using two of the nuclei as the foci of the ellipse with $\lambda \in [1,\infty)$  and  $\mu \in [-1,1]$ and are given by 
\begin{align}
\lambda &= \frac{r_{\text{a}}+r_{\text{b}} }{R_{\text{ab}}}, \\
\mu &= \frac{r_{\text{a}}-r_{\text{b}} }{R_{\text{ab}}},
\end{align}
with $r_{\text{a}},r_{\text{b}}$ the relative position between the electron and the two nuclei labelled as A and B, as shown in \fig{Fig:Molecule_microcanonical}. $R_{\text{ab}}$ is the distance between the two nuclei that are used to define the elliptical coordinates.  The coordinate $\phi$  is the angle between the projection of the position vector $\mathbf{r}_i $ of the bound electron $i$ on the $xy$ plane and the positive $x$ axis. The $z$ axis goes through the two nuclei that define the elliptical coordinates. Hence, the angle $\phi$ defines the rotation angle around the $z$ axis, for more details see \cite{Agapi_2016}. Transforming from Cartesian to elliptical coordinates, we find that the microcanonical distribution has the following form
\begin{widetext}
\begin{align}\label{eq:multie_micro2}
\begin{split}
&\rho\left(\lambda_5, \mu_5, \phi_5,\lambda_6, \mu_6, \phi_6\right)=\mathcal{N}'\left(\frac{R^{3}_{ab}}{8}\right)^{2} \prod_{i=5}^{6} \left(\lambda_i^2-\mu_i^2\right) \sqrt{2(E_i-W_i(\lambda_5, \mu_5, \phi_5,\lambda_6, \mu_6, \phi_6)} \delta\left(E_i+I_{p,2}\right).
\end{split}
\end{align}
\end{widetext}
As in our previous derivation of the microcanonical distribution for one bound electron in the presence of three nuclei \cite{Agapi_2016}, we find that the distribution $\rho$ in Eq. \eqref{eq:multie_micro2} becomes infinite only when an electron is placed on top of the third nucleus, labelled as C. Hence, as for our derivation in Ref. \cite{Agapi_2016}, to eliminate this singularity we introduce an additional transformation $t^{\gamma}_i=\mu_i-\mu_c$ where 
\begin{equation}
\mu_c = \dfrac{ R_{ac} - R_{bc} }{ R_{ab} },
\end{equation}
with $R_{ac}$ being the distance between nuclei A and C, and similarly for the other distances. We select the value $\gamma=3$ as the lowest one which eliminates the singularity mentioned above \cite{Agapi_2016}. The final form of the microcanonical distribution is 
\begin{align}\label{eq:microcanonical}
\begin{split}
&\tilde{\rho}\left(\lambda_5, t_5, \phi_5,\lambda_6, t_6, \phi_6\right)\\
&=\left\{\begin{array}{ll}
\prod_{i=5}^6 |t_i^{\gamma -1}|\left(\lambda_i^2-(t_i^\gamma + \mu_c)^2\right) \sqrt{P_i} & \text { for all }  P_i \geq 0 \\
0 & \text { for any } P_i<0,
\end{array}\right.
\end{split}
\end{align}
where
\begin{align}
\begin{split}
P_i &= 2\left[ -I_{p,2}-W_i\left(\lambda_5, t_5, \phi_5,\lambda_6, t_6, \phi_6\right) \right],
\end{split}
\end{align}
with 
\begin{widetext}
\begin{align}
&W_i =\frac{Q_iQ_1}{|\mathbf{r}_{1}-\mathbf{r}_{i}|}+\frac{Q_iQ_2}{|\mathbf{r}_{2}-\mathbf{r}_{i}|}+\frac{Q_iQ_3}{|\mathbf{r}_{3}-\mathbf{r}_{i}|}  + \sum_{n=1}^{N_{\text{c}}}\sum_{\substack{\;{j=N_{\text{c}}+1} \\ {j} \neq {i}}}^{N} c_{i,j}(t_0)C_{j,n}(\mathcal{E}_{j},|\mathbf{r}_{1}-\mathbf{r}_{j}|,...,|\mathbf{r}_{N_{\text{c}}}-\mathbf{r}_{j}|) V_{\text{eff}}(\zeta_{j,n}(t_0),|\mathbf{r}_{n}-\mathbf{r}_{i}|) \label{eq:potential_energy_of_electron_i_ell}\\ 
&\text{where,} \nonumber \\
&|\mathbf{r}_{1}-\mathbf{r}_{i}| = \frac{R_{ab}(\lambda_i + t_i^\gamma + \mu_c)}{2} \nonumber \\
&|\mathbf{r}_{2}-\mathbf{r}_{i}| = \frac{R_{ab}(\lambda_i - t_i^\gamma - \mu_c)}{2} \nonumber \\
&|\mathbf{r}_{3}-\mathbf{r}_{i}| = \frac{R_{ab}}{2}\left[\left(\lambda_i^2+(t_i^\gamma + \mu_c )^2-1\right)-\frac{4 z_c}{R_{a b}} \lambda_i (t_i^\gamma + \mu_c )+\frac{4\left(x_c^2 + z_c^2\right)}{R_{a b}^2} -\frac{4x_c cos(\phi_i)}{R_{ab}}\sqrt{(\lambda_i^2 -1)(1-(t_i^\gamma +\mu_c)^2)}\right]^{1 / 2} \label{eq:r3ri}
\end{align}
\end{widetext}
The parameters $x_c$, $z_c$  are given below
\begin{align}
x_c &= \pm \sqrt{R_{a c}^2-\left(\frac{R_{a c}^2-R_{b c}^2+R_{a b}^2}{2 R_{a b}}\right)^2},\\
z_c &= \frac{R_{a c}^2-R_{b c}^2}{2 R_{a b}}.
\end{align}
The new distribution $\tilde{\rho}$ goes to zero when one of the electrons is placed on top of the nucleus of the nucleus C, i.e., when $\lambda_i = \lambda_c= \dfrac{ R_{ac} + R_{bc} }{ R_{ab} }, t_i = 0$ and $\phi_i = 0, 2\pi$.

Next, we generate initial conditions for the linear triatomic molecule $\mathrm{HeH_2^{+}}$ assuming the nuclei A,B,C correspond to the Hydrogen atom on the left, to the Hydrogen atom in the middle, and the Helium atom, as shown in \fig{Fig:Molecule}. We now identify the range of values of $\lambda_i,t_i,\phi_i$ so that $P_i \geq 0$ for each bound electron $i$. We find that $t_i\in [t_{\text{min}},t_{\text{max}}]$, $\phi_i\in [0,2 \pi]$ for each electron, with $t_{\text{min}} =-(1+\mu_c)^{1/\gamma}$ and  $t_{\text{max}} =(1-\mu_c)^{1/\gamma}$. That is,  $P_i \geq 0$ is satisfied for the whole range of values of $\phi_i$ and $t_i.$ In addition, for $P_i \geq 0$ to be satisfied we find that $\lambda_i$ cannot be larger than $\lambda_{\text{max}},$ i.e.,  $\lambda_i\in [1,\lambda_{\text{max}}].$ The value $\lambda_{\text{max}}$ is the same for both bound electrons. For these range of values, then, we find the maximum value $\tilde{\rho}_{\text{max}}$ of the microcanonical distribution $\tilde{\rho}\left(\lambda_5, t_5, \phi_5,\lambda_6, t_6, \phi_6\right)$ given in \eq{eq:microcanonical}. Next, we generate the uniform random numbers $\lambda_i\in [1,\lambda_{\text{max}}]$, $t_i\in [t_{\text{min}},t_{\text{max}}]$, $\phi_i\in [0,2 \pi]$ for each electron, and $\chi \in [0,\tilde{\rho}_{\text{max}}]$. If  $\tilde{\rho}\left(\lambda_5, t_5, \phi_5,\lambda_6, t_6, \phi_6\right)>\chi$ then the generated values of $\lambda_i, t_i$ and $\phi_i$ are accepted as initial conditions, otherwise, they are rejected and the sampling process starts again.

Once we find the $\lambda_i, t_i$ and $\phi_i,$ we obtain the position vector $\mathbf{r}_{i} = (r_{x,i},r_{y,i},r_{z,i})$ and the momentum vector $\mathbf{p}_{i} = (p_{x,i},p_{y,i},p_{z,i})$ of each electron $i$ as follows
\begin{align}
r_{x,i} &= \frac{R_{ab} \cos \left( \phi_i \right)}{2} \sqrt{ \left( \lambda_i^2 - 1\right)\left[ 1 - \left(t_i^\gamma + \mu_c\right)^2\right]  }\\
r_{y,i} &= \frac{R_{ab} \sin \left( \phi_i \right)}{2} \sqrt{ \left( \lambda_i^2 - 1\right)\left[ 1 - \left(t_i^\gamma + \mu_c\right)^2\right]  }\\
r_{z,i} &= \frac{R_{ab} \lambda_i \left(t_i^\gamma + \mu_c\right)}{2} \\
p_{x,i} &= \sqrt{P_i} \cos \left( \phi_{\mathbf{p},i} \right) \sqrt{1 - \nu_{\mathbf{p},i}^2}\\
p_{y,i} &=\sqrt{P_i} \sin \left( \phi_{\mathbf{p},i} \right) \sqrt{1 - \nu_{\mathbf{p},i}^2}\\
p_{z,i} &=\sqrt{P_i} \nu_{\mathbf{p},i},
\end{align}
where $\phi_{\mathbf{p},i} \in [0,2 \pi]$ and $\nu_{\mathbf{p},i} \in [-1,1]$ define the momentum $\mathbf{p}_i$ in spherical coordinates. Following the above described formulation, we obtain the initial conditions of the electron with respect to the origin of the coordinate system. However, for our computations we need to obtain the initial conditions for the position of the electron with respect to the center of mass of the triatomic molecule. To do so, we shift the coordinates by $\mathbf{R_{\text{cm}}}=(X_{ \text{cm} }, 0 , Z_{ \text{cm} })$ with
\begin{align}
X_{ \text{cm}} &=  \frac{m_{c} x_{c}}{m_{a} + m_{b} + m_{c}} \label{eq:xcm}\\
Z_{ \text{cm}} &=  \frac{ (m_{b}-m_{a})R_{ab}/2 + m_{c} z_{c}}{m_{a} + m_{b} + m_{c}}. 
\end{align}
For  $\mathrm{HeH_2^{+}}$, $m_{a},m_{b}$ are the masses of the hydrogen atoms and $m_{c}$ the mass of the helium atom. We also note that since $\mathrm{HeH_2^{+}}$ is a linear molecule, $x_c$ is zero in Eqs. \eqref{eq:potential_energy_of_electron_i_ell}, \eqref{eq:r3ri} and \eqref{eq:xcm}. $Z_{ \text{cm}}$ can be seen in \fig{Fig:Molecule_microcanonical}. Also, we use the parameters, $I_{p,2}=-1.73$ a.u., $R_{ab} = |\mathbf{r}_{1}-\mathbf{r}_{2}|$ = 2.07 a.u., $R_{bc} = |\mathbf{r}_{2}-\mathbf{r}_{3}|$ = 2.06 a.u., $R_{ac} = |\mathbf{r}_{1}-\mathbf{r}_{3}|$ = 4.12 a.u., $Q_1=Q_2=1$ a.u. and $Q_3$ = 2 a.u. The distances and the second ionization energy of $\mathrm{HeH_2^{+}}$ were obtained using the quantum chemistry package MOLPRO \cite{MOLPRO_brief}, with the Hartree-Fock method employing the aug-cc-pV5Z basis set.

\section{Results}
Here, we employ a vector potential of the form
\begin{equation}\label{eq:vector_potential}
\mathbf{A}(\mathrm{y,t}) = -\frac{\mathrm{E}_0}{\omega}\exp \left[ - 2\ln (2)\left( \frac{\mathrm{c t - y}}{\mathrm{c} \tau} \right)^2 \right]   \sin ( \omega \mathrm{t}  - \mathrm{k y}) \hat{\mathbf{z}},
\end{equation}
where $\mathrm{k=\omega/c}$ is the  wave number of the laser field and  $\tau$ is the full width at half maximum of the pulse duration in intensity.  The direction of both the vector potential and the electric field is along the z axis. We take the propagation direction of the laser field to be along the y axis and hence the magnetic field points  along the x axis. The intensity of the field is $\mathrm{2 \times 10^{14} \; W/cm^2 }$ with a pulse duration in intensity of $\tau = 40$ fs at 800 nm. 

Using the ECBB  model for molecules, we focus on triple ionization (TI), frustrated triple ionization (FTI), double ionization (DI), and frustrated double ionization (FDI). Out of all events, we find that TI events account roughly for 1.2\% and FTI events with $n>2$ account for 0.3 \%, while DI and FDI with $n>2$ events account for 54\% and 9.5 \%, respectively. Hence, FDI is a major ionization process in strongly driven molecules. In triple ionization, three electrons escape and $\mathrm{He^{2+}}$ and two $\mathrm{H^+}$ ions are formed. In frustrated triple ionization,  two electrons escape and one electron finally remains  bound at a Rydberg state either at  $\mathrm{He^{2+}}$ or at  one of the two  $\mathrm{H^{+}}$ ions.  We also find that the formation of $\mathrm{He^{+*}+2H^{+}}$ is roughly 2.5 times more likely than the formation of  $\mathrm{He^{2+}+H^{+}+H^{*}}$.

 In double ionization, two electrons escape, while one remains bound.
For the vast majority of DI events, we find that the bound electron  has principal quantum number $n = 1$. Also,  it is three times more likely for the final fragments to be  $\mathrm{He^{+}+2H^{+}}$ rather than $\mathrm{He^{2+}+H^{+}+H}.$ That is, in DI, it is three times more likely for the electron to remain bound at He$^{2+}$. In frustrated double ionization,  an electron  escapes, an electron remains bound at an $n = 1$ state, and another electron remains bound at a Rydberg state. In FDI, we have several possibilities for the formation of different ions depending on which ions the bound electrons are attached to.  We find that 
it is roughly four times more likely for the deeply bound electron to remain bound at $\mathrm{He^{2+}}$ versus at $\mathrm{H^{+}}$. 

Here, in FTI and FDI we do not include the  Rydberg  $n=2$ states, since an electron from the $n=1$ state of $\mathrm{H^{+}}$ tunnels to the $n=2$ state of $\mathrm{He^{2+}}$ resulting in a  large number of $n = 2$ states. These states were also not included in  our previous work  on the  strongly driven heteronuclear molecules $\mathrm{HeH^{+}}$ \cite{Vila_2018} and $\mathrm{HeH_2^{+}}$ \cite{PhysRevA.103.043109}.

Finally, we identify the principal quantum number $n$ for each Rydberg electron by first calculating the classical principal quantum number
\begin{equation}
n_c = \frac{1}{\sqrt{2|\epsilon_i(t_f)|}}, 
\end{equation}
with $\epsilon_i(t_f)$ being the energy of a bound electron at the end of the time propagation. Then, we assign a quantum number $n$ so that the following criterion is satisfied \cite{Comtois_2005},
\begin{equation}
\left[  \left( n-1\right)\left( n-\frac{1}{2}\right)n \right]^{1/3} \leq n_c \leq \left[  n\left( n+\frac{1}{2}\right)\left( n+1\right) \right]^{1/3}.
\end{equation}

\subsection{KER distributions}

\begin{figure}[H]
\centering
\includegraphics[width=\linewidth]{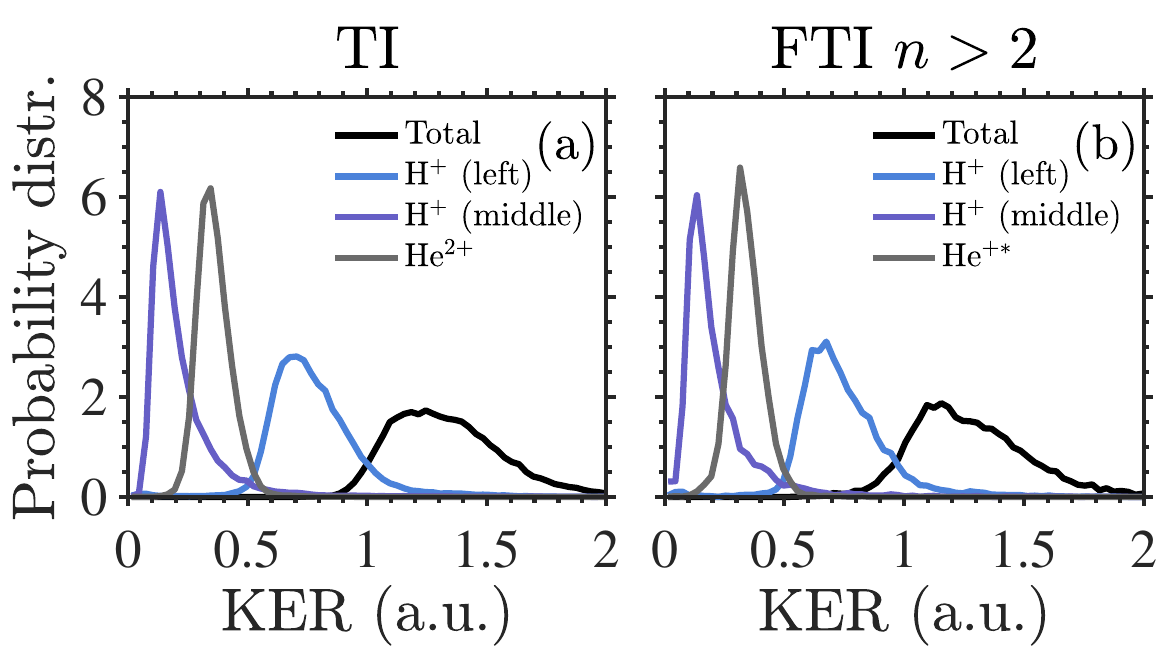}
\caption{Distribution of the sum of the final kinetic energies (black solid lines) of the ions produced in (a) triple ionization, and (b) frustrated triple ionization. The grey lines depict the distribution of the final kinetic energy of the $\mathrm{He^{2+}}$ ion  for TI, and  $\mathrm{He^{+*}}$ for FTI. The purple (light blue) lines depict the distribution of the final kinetic energy of the middle (left) $\mathrm{H^{+}}$ ion for TI and FTI. All distributions are normalized to one.}
\label{Fig:KER_TI_and_FTI}
\end{figure}

In \fig{Fig:KER_TI_and_FTI}, we plot the kinetic energy release (KER) of the final ion fragments for triple and the most probable route to frustrated triple ionization. As expected, we find that the  KER for TI and FTI are very similar. This is consistent with  the Rydberg electron in FTI remaining bound in a highly excited state. Hence, the Rydberg electron does not significantly screen  the core  it remains bound to. Given the similarity of the KER distributions for FTI and TI, for simplicity,  we next focus on describing the features of the KER for TI.  As in our previous work \cite{PhysRevA.103.043109}, we find that the left $\mathrm{H^{+}}$ ion is the fastest one, followed by $\mathrm{He^{+}}$ and the middle $\mathrm{H^{+}}$ ion. This is consistent with both Coulomb forces exerted on the left $\mathrm{H^{+}}$ ion being along the $-z$ axis. Similarly, both repulsive forces acting on $\mathrm{He^{+}}$ are along the $+z$ axis. However, the mass of He is four time larger than the mass of H. This results in He$^{+}$ having a smaller acceleration and hence kinetic energy  compared to the left $\mathrm{H^{+}}$ ion. Also, the middle $\mathrm{H^{+}}$ ion experiences repulsive forces from the left $\mathrm{H^{+}}$ ion and  from the $\mathrm{He^{+}}$ ion  in opposite directions. This small net Coulomb force on the middle $\mathrm{H^{+}}$ ion results  in its kinetic energy being smaller compared to  the other two ions.

\begin{figure}[H]
\centering
\includegraphics[width=\linewidth]{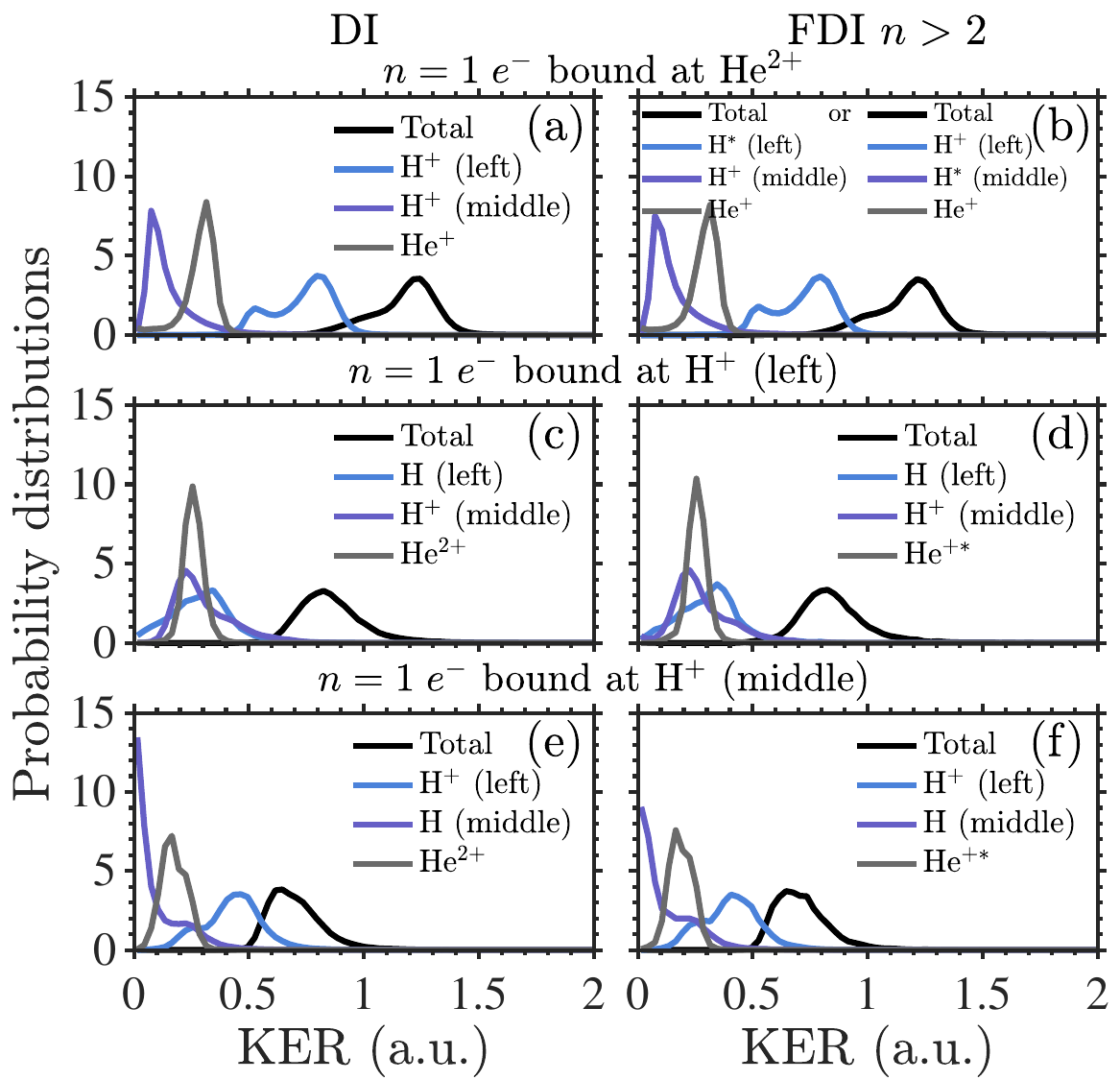}
\caption{Distribution of the sum of the final kinetic energies (black solid lines) of the ions produced in (a),(c),(e) double ionization, and (b),(d),(f) frustrated double ionization with the $n=1$ electron bound at $\mathrm{He^{2+}}$ (top row), bound at the left $\mathrm{H^{+}}$ ion (middle row) and bound at the middle $\mathrm{H^{+}}$ ion (bottom row). The grey lines depict the distribution of the final kinetic energy of the $\mathrm{He^{+}}$ ion  for DI and FDI (top row), and $\mathrm{He^{2+}}$ for DI and $\mathrm{He^{+*}}$ for FDI (middle and bottom row). The purple (light blue) lines depict the distribution of the final kinetic energy of the middle (left) $\mathrm{H^{+}}$ ion fragment for DI and $\mathrm{H^{*}}$ or $\mathrm{H^{+}}$ for FDI (top row),  $\mathrm{H^{+} \; (H)}$ ion  for DI and FDI (middle row), and $\mathrm{H \; (H^{+})}$ ion  for DI and FDI (bottom row). All distributions are normalized to one.}
\label{Fig:KER_DI_and_FDI}
\end{figure}
In \fig{Fig:KER_DI_and_FDI}, we plot the kinetic energy release of the final ion fragments for double and frustrated double ionization. In what follows, we focus on the most probable channels of FDI, namely, the $\mathrm{He^{+} + H^{+} + H^{*}}$ [see \fig{Fig:KER_DI_and_FDI}(b)] which accounts for 65\% of FDI and the channel $\mathrm{He^{+*} + H^{+} + H}$  [see Figs. \ref{Fig:KER_DI_and_FDI}(d) and \ref{Fig:KER_DI_and_FDI}(f)] which account for 17\% of FDI. We plot the KER when the bound $n=1$ electron is attached to the $\mathrm{He^{2+}}$ ion (top row),    to the left $\mathrm{H^{+}}$ ion (middle row), and to the middle $\mathrm{H^{+}}$ ion (bottom row). As for TI and FTI, we find that the KER distributions for DI and FDI are very similar. Hence, for simplicity, we next focus on describing the features of the KER distribution for DI. When the deeply bound electron is attached to $\mathrm{He^{2+}}$ (top row), we find that the left $\mathrm{H^{+}}$ is the fastest ion, followed by $\mathrm{He^{+}}$ and the middle $\mathrm{H^{+}}$ ion. Indeed, the Coulomb repulsive forces on  the left $\mathrm{H^{+}}$ ion from the other two ions  are both along the $-z$ axis, on $\mathrm{He^{+}}$ both forces are along the $+z$ axis, and on the middle $\mathrm{H^{+}}$ ion the two forces are  in opposite directions. When the deeply bound electron is attached to the left $\mathrm{H^{+}}$ ion (middle row), the electron screens the charge of the core, resulting in a smaller Coulomb repulsion between the left $\mathrm{H}$ fragment and $\mathrm{He^{2+}}$. Hence, the kinetic energy  of each of the two fragments is smaller compared to their kinetic energy when the deeply bound  electron is attached to $\mathrm{He^{+}}$. The reduction in kinetic energy is larger for the left $\mathrm{H}$ fragment since both the Coulomb forces from the other two ions are now smaller.  This is clearly seen by comparing the grey and light blue lines in \fig{Fig:KER_DI_and_FDI}(b) with the ones in  \fig{Fig:KER_DI_and_FDI}(a). In contrast, the total Coulomb repulsion on the middle $\mathrm{H^{+}}$ ion is increased. Indeed,  the repulsion from the left $\mathrm{H}$ fragment towards the $z$ axis is decreased while the repulsion from $\mathrm{He^{2+}}$ towards the $-z$ axis is increased (the $n=1$ electron is no longer attached to  $\mathrm{He^{2+}}$). Hence, the kinetic energy of the middle  $\mathrm{H^{+}}$ is increased,   compare the purple  line in \fig{Fig:KER_DI_and_FDI}(b) with the one in \fig{Fig:KER_DI_and_FDI}(a).  When the deeply bound electron is attached  to the middle $\mathrm{H^{+}}$ ion, the repulsive forces on this fragment are smaller resulting in its smaller kinetic energy, compare  the purple  line in \fig{Fig:KER_DI_and_FDI}(e) with the one in  \fig{Fig:KER_DI_and_FDI}(c). Also, the kinetic energy of the left $\mathrm{H^{+}}$ ion increases since the  deeply bound electron now screens the middle  $\mathrm{H}$ fragment, compare  the light blue  line in 
\fig{Fig:KER_DI_and_FDI}(e) with the one in  \fig{Fig:KER_DI_and_FDI}(c). Finally, the kinetic energy  of $\mathrm{He^{2+}}$ is smaller since the deeply bound electron is attached to the H$^{+}$ ion that is closer to  He$^{2+}$,  compare  the grey   lines in \fig{Fig:KER_DI_and_FDI}(e) and  \fig{Fig:KER_DI_and_FDI}(c)

\begin{figure}[H]
\centering
\includegraphics[width=\linewidth]{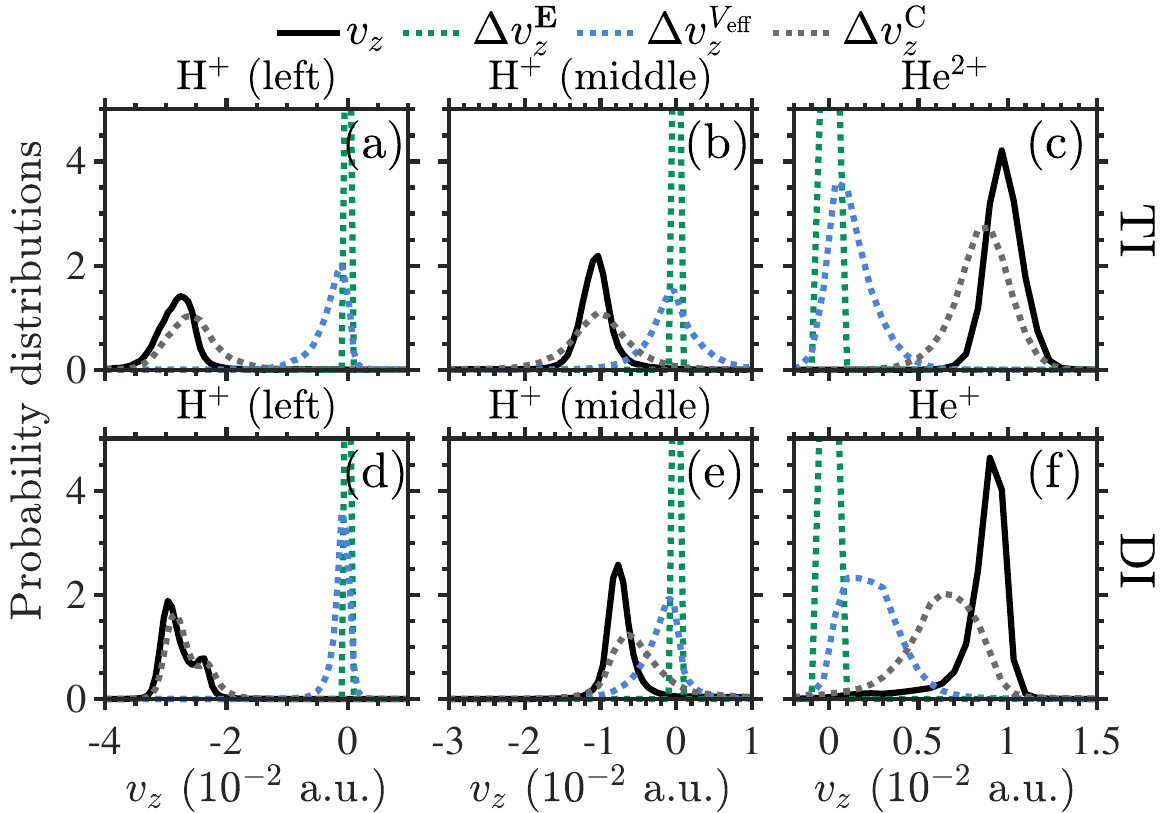}
\caption{Distribution of the $z$ component of the final velocity of the ions, $v_z$, (solid black lines)  and of the change of $v_{z}$ in the time interval $[t_0,t_f]$ due to the forces from the electric field $\Delta v_{z}^{\mathbf{E}}$ (dotted green lines), from the effective potential $\Delta v_{z}^{V_{\text{eff}}}$ (dotted light blue lines), and from the Coulomb potential $\Delta v_{z}^{C}$ (dotted dark grey lines). The top row corresponds to TI  and the bottom row to DI  with the $n=1$ electron bound at $\mathrm{He^{2+}}$. }
\label{Fig:Vz_contributions}
\end{figure}

Moreover, we find that the KER distribution of the left $\mathrm{H^{+}}$ ion has a pronounced double peak structure for DI and FDI when the deeply bound electron is attached to $\mathrm{He^{2+}}$ [see light blue lines in Figs. \ref{Fig:KER_DI_and_FDI}(a) and \ref{Fig:KER_DI_and_FDI}(b)]. 
To identify the origin of this double peak, it suffices to focus on the DI process.
 In \fig{Fig:Vz_contributions}, we plot the final $z$ component of the velocity of the ions along the electric field, $v_z$, for TI (top row) and DI (bottom row). Also, we plot the contribution  to this velocity from the electric field, $\Delta v_{z}^{\mathbf{E}}$,  and from the forces due to the Coulomb,  $\Delta v_{z}^{C}$, and  the effective,  $\Delta v_{z}^{V_{\text{eff}}}$, potentials.  Both for TI and DI, we find that the velocity $v_z$ of the left H$^{+}$ ion is mostly determined by the Coulomb forces acting on this ion. For  DI, the contribution of the Coulomb forces to the velocity of the left H$^{+}$ ion has a clear double peak structure, see \fig{Fig:Vz_contributions}(d). This gives rise to the double peak structure in the kinetic energy of the left H$^{+}$ ion in \fig{Fig:KER_DI_and_FDI}(a).     In contrast, the velocities $v_z$ of the middle H$^{+}$ and of the He$^{2+}$ (He$^{+}$) ion for TI (DI)  are determined by the forces from both the Coulomb and the effective potentials. The role of the effective potential is more pronounced for the DI process, consistent with one electron remaining bound.

\begin{figure}[H]
\centering
\includegraphics[width=\linewidth]{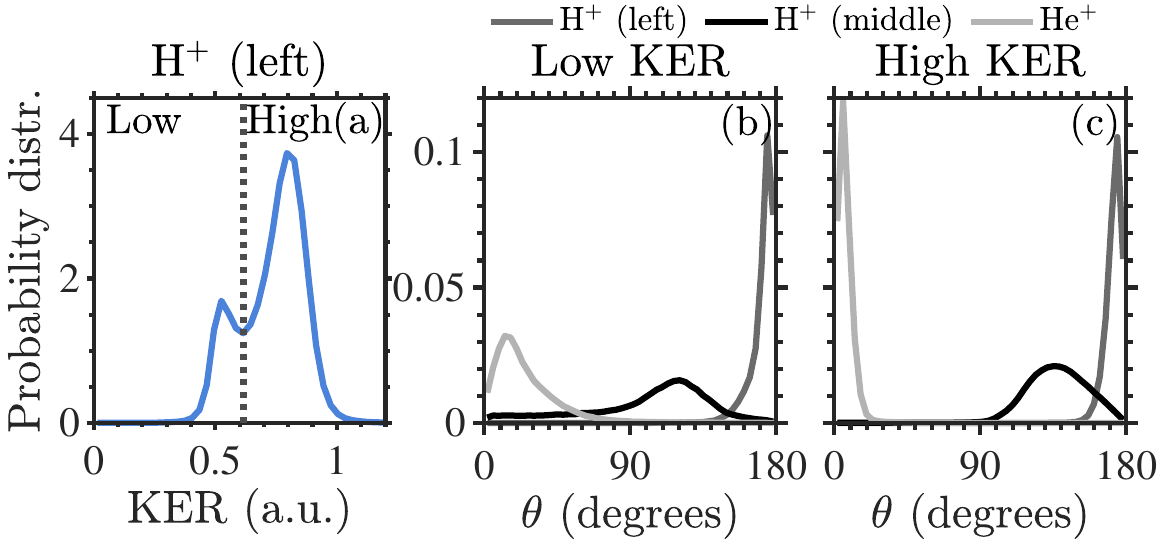}
\caption{(a) Distribution of the final kinetic energy of the left $\mathrm{H^{+}}$ ion for DI when the $n=1$ electron is bound at $\mathrm{He^{2+}}$. Angular distributions of the left $\mathrm{H^{+}}$,  middle $\mathrm{H^{+}}$ and $\mathrm{He^{+}}$ ions for DI  when the kinetic energy of the  left $\mathrm{H^{+}}$ is low (b) and high (c). }
\label{Fig:Angles_DI_low_high}
\end{figure}

\begin{figure}[H]
\centering
\includegraphics[width=\linewidth]{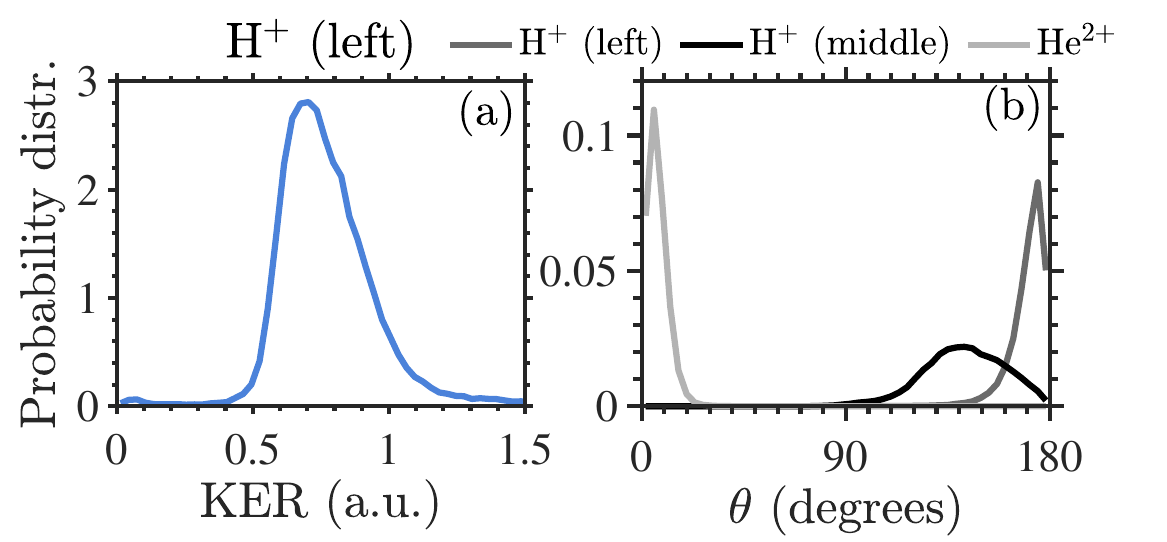}
\caption{(a) Distribution of the final kinetic energy of the left $\mathrm{H^{+}}$ ion for TI. (b) Angular distributions of the left $\mathrm{H^{+}}$,  middle $\mathrm{H^{+}}$ and $\mathrm{He^{2+}}$ ions for TI. }
\label{Fig:Angles_TI}
\end{figure}

Next, for DI, we show how the Coulomb forces exerted on the left H$^{+}$ ion, when the deeply bound electron is attached at He$^{2+}$, result in a low and high energy peak in the KER of  the left H$^{+}$ ion, see light blue line in \fig{Fig:KER_DI_and_FDI}(a). To do so, we plot the angle of escape of each ion  with respect to the $z$ axis. The angular distributions that correspond to the low and high energy peaks [see  \fig{Fig:Angles_DI_low_high}(a)] are shown in   \fig{Fig:Angles_DI_low_high}(b) and  \fig{Fig:Angles_DI_low_high}(c), respectively. 
  We find that  the two peaks in the kinetic energy distribution of the left   H$^{+}$ ion are associated with a different range of angles of escape  of the middle  $\mathrm{H^{+}}$ as well as of the He$^{+}$ ions. For both peaks, the left $\mathrm{H^{+}}$ ion escapes along the $-z$ axis, see dark grey lines in Figs. \ref{Fig:Angles_DI_low_high}(b)-(c). Moreover, all ions have roughly the same charge and the middle  $\mathrm{H^{+}}$ ion is closer compared to $\mathrm{He^{+}}$ to the left $\mathrm{H^{+}}$ ion. Hence, it is the Coulomb repulsion between the two $\mathrm{H^{+}}$ ions that mostly determines the final kinetic energy of the left $\mathrm{H^{+}}$ ion. We find that for the lower energy peak  the middle $\mathrm{H^{+}}$ ion escapes with a very wide range of angles away from the  -$z$ axis, compared to a much smaller range for the high energy peak.  When the middle $\mathrm{H^{+}}$ ion escapes with larger angles with respect to the -$z$ axis and, hence, with respect to the left H$^{+}$, the Coulomb repulsion between the two $\mathrm{H^{+}}$ ions is smaller resulting in a smaller kinetic energy of the left $\mathrm{H^{+}}$ ion. 
   In \fig{Fig:Angles_TI}(b), we show that for TI  the distributions of the angles of escape of the ions are very similar to the angles  corresponding to the high energy peak for DI, compare  \fig{Fig:Angles_TI}(b) with \fig{Fig:Angles_DI_low_high}(c). As a result,  the kinetic energy distribution  of the left H$^{+}$  ion for TI is similar to the part of the  distribution for DI that corresponds to the high energy peak.
  
Finally, a comparison of the KER for TI, FTI and DI obtained with the ECBB model and with the previous model in Ref. \cite{PhysRevA.103.043109} reveals that the KER have larger values  for the ECBB model. This is consistent with   taking into account the repulsion between the bound electrons using effective potentials in the ECBB model. In our previous more primitive model, the repulsion between bound electrons is turned off. Due to this repulsion via the effective potentials the electrons are less bound to the nuclei they are attached to. Hence, the electrons screen the nuclei less, leading to higher Coulomb repulsion between the nuclei and, therefore, to larger values of the KER. 
Other differences in the KER of the left and middle $\mathrm{H^{+}}$ ions when using the ECBB model versus our previous model in  Ref. \cite{PhysRevA.103.043109} are due to  the effective potentials in the ECBB model significantly influencing the final velocities of the middle $\mathrm{H^{+}}$ and He$^{+}$ ions, see \fig{Fig:Vz_contributions}.

\subsection{Correlation in electron escape}

\begin{figure}[t]
\centering
\includegraphics[width=\linewidth]{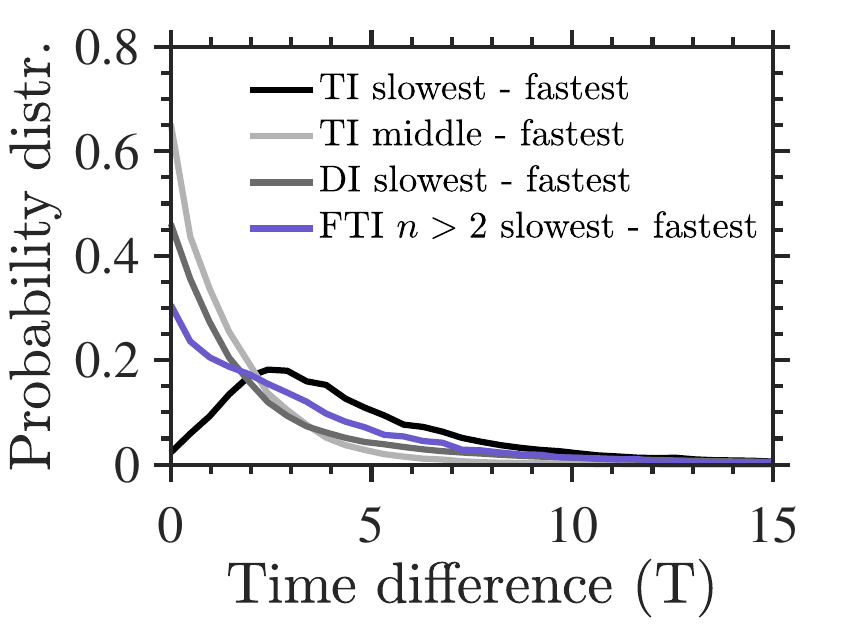}
\caption{Distribution of the  difference of the ionization times between the fastest and second fastest electrons as well as the fastest and slowest electrons in TI and between the fastest and slowest electrons in FTI and DI.}
\label{Fig:Ion_times_differences}
\end{figure}

In \fig{Fig:Ion_times_differences}, we plot the distribution of the difference of the ionization times between the fastest and second fastest electrons as well as the fastest and slowest electrons in TI and between the fastest and slowest electrons in FTI and DI. We find that the electron that ionizes second has a significant probability to do so with a small time difference from the fastest one, with the time difference being the smallest in TI, followed by DI and then by FTI. The distributions in all three processes extend up to 10 periods (T) of the laser field. In contrast, compared to the fastest electron, the time the last electron ionizes in TI has a distribution that peaks roughly around three periods of the laser field. This suggests that the last to ionize electron escapes mainly due to enhanced ionization and not due to a recollision, i.e. the electronic correlation is weak. This is consistent with HeH$_{2}^{+}$ being driven  by  a long and intense laser pulse in the current study. 

A comparison between the distributions of the difference of the ionization times for TI, FTI and DI of HeH$_{2}^{+}$  obtained with  the ECBB model versus its predecessor  model in Ref. \cite{PhysRevA.103.043109}, reveals that these distributions for the ECBB model peak at smaller times and are less wide. This is consistent with the interaction between bound electrons being accounted for via effective potentials in the ECBB model. 
As a result, following a return of an electron to the core, energy transfer between bound electrons can take place, leading to possible ionization or excitation. This in turn leads to electrons escaping faster which is consistent with the KER of the fragments having larger values for the ECBB model, as discussed above.

\subsection{$n$ quantum numbers}

\begin{figure}[t]
\centering
\includegraphics[width=\linewidth]{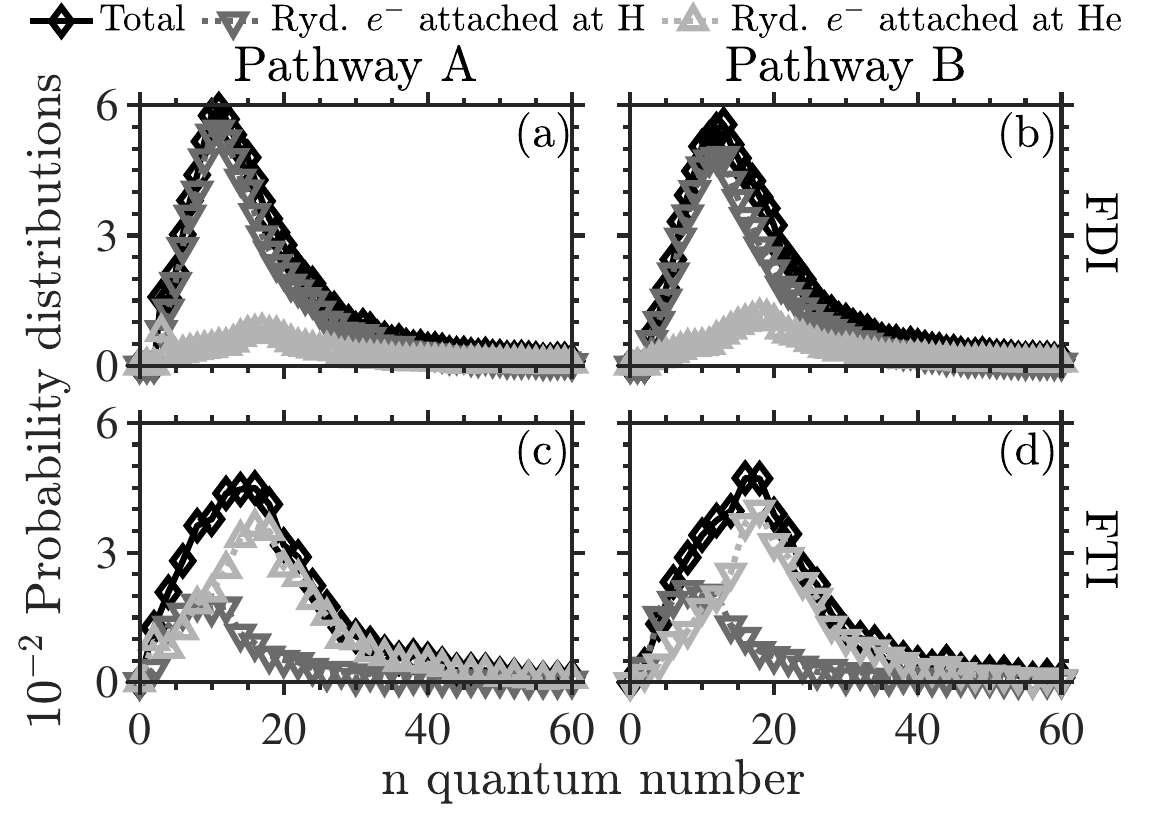}
\caption{Distribution of the principal n quantum number for pathways A (a) and B (b) of FDI and for pathways A (c) and B (d) of FTI. The distribution of the n quantum number is also plotted separately when the Rydberg electron remains attached to $\mathrm{He^{2+}}$ for FTI and He$^{+}$ for FDI (light gray lines) and when it remains attached to $\mathrm{H^{+}}$ (dark gray lines).}
\label{Fig:Quantum_numbers}
\end{figure}

Next, we investigate the distribution of the principal $n$ quantum number of the two main pathways A and B of FTI and FDI, see  \fig{Fig:Quantum_numbers}. We find that FDI with $n>2$ is a major ionization process accounting for roughly 9.5\% of all events while  FTI is an order of magnitude less probable.
 Also, we find that pathways A (54 \%) and B  (46 \%) with $n > 2$ contribute roughly the same to FTI, while for FDI, pathway B contributes significantly more (70\%) than pathway A. 
  This can be understood in terms of electronic correlation.
  FTI, most likely, occurs when the slowest electron  finally remains bound in an excited state. As for the slowest electron in TI, see black line in  \fig{Fig:Ion_times_differences}, the slowest electron in FTI can gain energy  both from  the initially tunneling electron returning to the molecular ion as well as from an enhanced ionization process. Hence, when this electron remains bound in a Rydberg state, it does so either through pathway B, related to energy gain from the returning electron, and through pathway A, related to energy gain from an enhanced ionization process \cite{Emmanouilidou_2012}. FDI, most likely, occurs when the slowest out of the two electrons that ionize in DI finally remains in an excited state. However, the slowest electron in DI mainly gains energy from the  electron returning to the molecular ion, associated with pathway B of FDI. This is supported by the significantly  faster ionization time of the second electron  in DI and TI (dark and light grey lines  in  \fig{Fig:Ion_times_differences}) compared to the ionization times of the slowest electron in TI  (black line in  \fig{Fig:Ion_times_differences}).
 
 For FDI, we find that it  is  significantly more likely for the Rydberg electron to be attached to the H$^{+}$ ion versus the He$^{2+}$ ion.  Indeed,  in  \fig{Fig:Quantum_numbers}(a) and 
\fig{Fig:Quantum_numbers}(b), it is clearly seen that   the probability for the Rydberg electron to be attached to He$^{2+}$ (area under the light grey lines) is much smaller than the probability to be attached to one of the  H$^{+}$ ions (area under the dark grey lines). This is consistent with 65\% of FDI events having the Rydberg electron attached to one of the H$^{+}$ ions, while the deeply bound electron is attached to the He$^{2+}$, see \fig{Fig:KER_DI_and_FDI}(b). 
Only 21\% of FDI events have the Rydberg electron attached to He$^{2+}$, while the  deeply bound electron is attached to one of the H$^{+}$ ions, see \fig{Fig:KER_DI_and_FDI}(d) and \fig{Fig:KER_DI_and_FDI}(f). The significantly higher probability for the Rydberg electron to be attached to one of the H$^{+}$ ions,  is consistent with the bound $n=1$ electron staying mostly attached to the He$^{2+}$ ion both for DI and FDI. In this case, all nuclei have roughly charge of one. In addition, the bound $n=1$ electron repels the Rydberg electron from the He$^{+}$ ion, resulting in the Rydberg electron   being more likely to stay bound in one of the two H$^{+}$ ions. The Rydberg electron can also  remain bound at He$^{+}$, a less likely process that  we do not show in  Figs. \ref{Fig:KER_DI_and_FDI} and \ref{Fig:Quantum_numbers}.

For FTI, in contrast to FDI, it is roughly 2.5 times more likely for the Rydberg electron to remain attached to He$^{2+}$ versus the H$^{+}$ ions, compare the area under the light and dark grey lines in Figs. \ref{Fig:Quantum_numbers}(c) and \ref{Fig:Quantum_numbers}(d). This is consistent with an  electron being significantly more likely to be attracted and remain bound at He$^{2+}$ with charge two versus at H$^{+}$ with charge one.

Also, for FDI and FTI, we find that for both pathways the distribution of the $n$ quantum number peaks around $n=18$  when the Rydberg electron is attached to He$^{2+}$ compared to the significantly smaller $n$ values when the Rydberg electron is attached to H$^{+}$.This comes as no surprise. Indeed, we assume that the electron that tunnel ionizes last and remains bound in a Rydberg state has roughly the same energy for attachment at  He$^{2+}$ or at H$^{+}$. Given that He$^{2+}$ has twice the charge   of H$^{+}$, it follows that a Rydberg  $n$ state  of H$^{+}$ corresponds to roughly a  Rydberg $2n$ state  of  He$^{2+}$. In addition, we find that the distribution of the $n$ number peaks at higher values for pathway A of FDI versus FTI. This is consistent with an $n=1$ electron remaining bound in FDI resulting in a higher screening of the cores  in FDI compared to FTI and hence higher energies of the Rydberg electron for FDI.

\section{Conclusions}
We have developed a general three-dimensional semiclassical model   for the study of correlated multielectron escape during fragmentation of molecules driven by intense infrared laser pulses. This model fully accounts for the motion of all electrons and nuclei. Moreover, it is developed in the nondipole approximation, fully accounting for the magnetic field of the laser pulse. This model is a generalization of a model we have previously developed for atoms. In this model, referred to as ECBB model for molecules, the interaction of each quasifree or bound electron with the cores and each quasifree electron with a bound electron is treated exactly, fully accounting for the Coulomb singularity. To avoid artificial autoionization, the interaction of a pair of bound electrons is treated through effective Coulomb potentials. We employ the ECBB model in the context of the linear triatomic HeH$_{2}^{+}$ molecule. We focus our studies on triple and double as well as frustrated triple and double ionzation. We find that the sum of the final kinetic energies of all ion fragments are larger when described by the ECBB model versus a predecessor of the ECBB model that does not account for the interaction between bound electrons. This suggests that the interaction between bound electrons allows for a more correlated electron-electron escape which occurs faster, leading to a Coulomb explosion of the nuclei at shorter distances. Finally, the ECBB model allows for the study of frustrated double ionization, a major ionization process. This process was  not accessible with our previous models, since it has two bound electrons and the interaction between bound electrons was not accounted for. 
We expect that the ECBB model for strongly driven molecules will pave the way for previously inaccessible studies of multielectron ionization processes during fragmentation of strongly driven molecules.

\section{Acknowledgements}
A.E. and G.P.K. acknowledge EPSRC Grant No. EP/W005352/1. The authors acknowledge the use of the UCL Myriad High Performance Computing Facility (Myriad@UCL), the use of the UCL Kathleen High Performance Computing Facility (Kathleen@UCL), and associated support services in the completion of this work.

\vspace{1cm}
\appendix
\section{Derivation of  terms in the chain rule. }\label{Appendix_A}
We find that the terms in the chain rule in \eqref{eq:total_time_derivative} are given by

{\allowdisplaybreaks
\begin{widetext}
\begin{align}
&{\frac{\partial \left[\mathcal{E}_{j}(t) - H\right]}{\partial \mathbf{r}_{{j}}}\cdot\dot{\mathbf{r}}_{{j}} } ={\frac{\partial \left[ - Q_j{\mathbf{r}_{j} \cdot \mathbf{E}\left(\mathbf{r}_{j}, t\right)}-\sum_{{i}=N_{\text{c}}+1}^{{N-1}}\sum_{{m}={i}+1}^{{N}} \left[ {1-c_{i,m}(t)}\right]\frac{{Q_i}{Q_m}}{|\mathbf{r}_i-\mathbf{r}_m|} \right]}{\partial \mathbf{r}_{{j}}}\cdot\dot{\mathbf{r}}_{{j}}} \nonumber \\
&- { \frac{\left[ \sum_{\substack{\;{i=N_{\text{c}}+1} \\ {i} \neq {j}}}^{{N}}\sum_{n=1}^{N_{\text{c}}} c_{i,j}(t) C_{j,n}(\mathcal{E}_{j},|\mathbf{r}_{1}-\mathbf{r}_{j}|,...,|\mathbf{r}_{N_{\text{c}}}-\mathbf{r}_{j}|)V_{\text{eff}}(\zeta_{j,n},|\mathbf{r}_{n}-\mathbf{r}_{i}|) \right]}{\partial \mathbf{r}_j}\cdot \mathbf{\dot{r}}_{j}} \nonumber \\
&={-Q_j \mathbf{E}\left(\mathbf{r}_{j}, t\right)\cdot  \mathbf{\dot{r}}_{j}  + \sum_{{i}={N_{\text{c}}}+1}^{{N-1}}\sum_{{m}={i}+1}^{{N}}[1-c_{i,m}(t)] {\frac{Q_i Q_m(\mathbf{r}_{i}-\mathbf{r}_{n})}{|\mathbf{r}_{i}-\mathbf{r}_{m}|^3}\left(\delta_{i,j} - \delta_{m,j}\right)  \cdot \dot{\mathbf{r}}_{{j}}  }   } \nonumber \\
&- { Q_j\mathbf{r}_{j} \cdot \left[ \frac{\partial \mathbf{E}\left(\mathbf{r}_{j}, t\right)}{\partial \mathbf{r}_{j}} \cdot \mathbf{\dot{r}}_{j}\right]  - {\sum_{\substack{\;{i={N_{\text{c}}}+1} \\ {i} \neq {j}}}^{{N}}}{\sum_{n=1}^{N_{\text{c}}} c_{i,j}(t) \frac{\partial C_{j,n}(\mathcal{E}_{j},|\mathbf{r}_{1}-\mathbf{r}_{j}|,...,|\mathbf{r}_{N_{\text{c}}}-\mathbf{r}_{j}|)}{\partial \mathbf{r}_j}V_{\text{eff}}(\zeta_{j,n},|\mathbf{r}_{n}-\mathbf{r}_{i}|)\cdot \mathbf{\dot{r}}_{j}}}\\
&{\sum_{n=1}^{N_{\text{c}}}\frac{\partial \mathcal{E}_{j}(t)}{\partial \mathbf{r}_{{n}}}\cdot\dot{\mathbf{r}}_{{n} }}= {\sum_{n=1}^{N_{\text{c}}}{\sum_{\substack{\;{i={N_{\text{c}}}+1} \\ {i} \neq {j}}}^{{N}}} c_{i,j}(t)\sum_{b=1}^{N_{\text{c}}}\left[ \delta_{n,b}C_{i,b}\frac{\partial V_{\text{eff}}(\zeta_{i,b},|\mathbf{r}_{b}-\mathbf{r}_{j}|)}{\partial \mathbf{r}_n}+ V_{\text{eff}}(\zeta_{i,b},|\mathbf{r}_{b}-\mathbf{r}_{j}|)\frac{\partial C_{i,b}}{\partial \mathbf{r}_n}\right]  \cdot \dot{\mathbf{r}}_{{n}} }  \nonumber\\
&+{\sum_{n=1}^{N_{\text{c}}}\sum_{b=1}^{N_{\text{c}}}\left[-\delta_{n,b}\frac{Q_b Q_j(\mathbf{r}_{b}-\mathbf{r}_{j})}{|\mathbf{r}_{b}-\mathbf{r}_{j}|^3} \right] \cdot \dot{\mathbf{r}}_{{n}}}\nonumber\\
&= {\sum_{n=1}^{N_{\text{c}}}{\sum_{\substack{\;{i={N_{\text{c}}}+1} \\ {i} \neq {j}}}^{{N}}} c_{i,j}(t)\left[C_{i,n}(\mathcal{E}_{i},|\mathbf{r}_{1}-\mathbf{r}_{i}|,...,|\mathbf{r}_{N_{\text{c}}}-\mathbf{r}_{i}|)\frac{\partial V_{\text{eff}}(\zeta_{i,n},|\mathbf{r}_{n}-\mathbf{r}_{j}|)}{\partial \mathbf{r}_n}+\sum_{b=1}^{N_{\text{c}}} V_{\text{eff}}(\zeta_{i,b},|\mathbf{r}_{b}-\mathbf{r}_{j}|)\frac{\partial C_{i,b}}{\partial \mathbf{r}_n}\right]  \cdot \dot{\mathbf{r}}_{{n}} } \nonumber \\
&+{\sum_{n=1}^{N_{\text{c}}}\left[-\frac{Q_n Q_j(\mathbf{r}_{n}-\mathbf{r}_{j})}{|\mathbf{r}_{n}-\mathbf{r}_{j}|^3} \right] \cdot \dot{\mathbf{r}}_{{n}}}\\
&{\sum_{\substack{\;{i={N_{\text{c}}}+1} \\ {i} \neq {j}}}^{{N}}}{\frac{\partial \mathcal{E}_{j}(t)}{\partial \mathbf{r}_{{i}}}\cdot\dot{\mathbf{r}}_{{i} }}
= {{\sum_{\substack{\;{i={N_{\text{c}}}+1} \\ {i} \neq {j}}}^{{N}}}{\sum_{\substack{\;{l={N_{\text{c}}}+1} \\ {l} \neq {j}}}^{{N}}} \sum_{n=1}^{N_{\text{c}}}c_{l,j}(t)\left[ V_{\text{eff}}(\zeta_{l,n},|\mathbf{r}_{n}-\mathbf{r}_{j}|)\delta_{i,l}\frac{\partial C_{l,n}(\mathcal{E}_{l},|\mathbf{r}_{1}-\mathbf{r}_{l}|,...,|\mathbf{r}_{N_{\text{c}}}-\mathbf{r}_{l}|)}{\partial \mathbf{r}_i}\right]  \cdot \dot{\mathbf{r}}_{{i}} } \nonumber\\
&= {{\sum_{\substack{\;{i={N_{\text{c}}}+1} \\ {i} \neq {j}}}^{{N}}} \sum_{n=1}^{N_{\text{c}}}c_{i,j}(t)\left[V_{\text{eff}}(\zeta_{i,n},|\mathbf{r}_{n}-\mathbf{r}_{j}|)\frac{\partial C_{i,n}(\mathcal{E}_{i},|\mathbf{r}_{1}-\mathbf{r}_{i}|,...,|\mathbf{r}_{N_{\text{c}}}-\mathbf{r}_{i}|)}{\partial \mathbf{r}_i}\right]  \cdot \dot{\mathbf{r}}_{{i}} } \\
&{{\sum_{\substack{\;{i={N_{\text{c}}}+1} \\ {i} \neq {j}}}^{{N}}}\frac{\partial \mathcal{E}_{j}(t)}{\partial \mathcal{E}_i} \dot{\mathcal{E}_i}} = {\sum_{\substack{\;{i={N_{\text{c}}}+1} \\ {i} \neq {j}}}^{{N}}} \dot{\mathcal{E}_i}{\sum_{\substack{\;{l={N_{\text{c}}}+1} \\ {l} \neq {j}}}^{{N}}}\sum_{n=1}^{N_{\text{c}}}c_{l,j}(t)\delta_{i,l}\left[ C_{l,n}(\mathcal{E}_{l},|\mathbf{r}_{1}-\mathbf{r}_{l}|,...,|\mathbf{r}_{N_{\text{c}}}-\mathbf{r}_{l}|)\frac{\partial V_{\text{eff}}(\zeta_{l,n},|\mathbf{r}_{n}-\mathbf{r}_{j}|)}{\partial \mathcal{E}_i}\right. \nonumber \\
& \left.+V_{\text{eff}}(\zeta_{l,n},|\mathbf{r}_{n}-\mathbf{r}_{j}|)\frac{\partial C_{l,n}(\mathcal{E}_{l},|\mathbf{r}_{1}-\mathbf{r}_{l}|,...,|\mathbf{r}_{N_{\text{c}}}-\mathbf{r}_{l}|)}{\partial \mathcal{E}_i}\right] \nonumber  \\
&= {\sum_{\substack{\;{i={N_{\text{c}}}+1} \\ {i} \neq {j}}}^{{N}}} \dot{\mathcal{E}_i}\sum_{n=1}^{N_{\text{c}}}c_{i,j}(t)\left[ C_{i,n}(\mathcal{E}_{i},|\mathbf{r}_{1}-\mathbf{r}_{i}|,...,|\mathbf{r}_{N_{\text{c}}}-\mathbf{r}_{i}|)\frac{\partial V_{\text{eff}}(\zeta_{i,n},|\mathbf{r}_{n}-\mathbf{r}_{j}|)}{\partial \mathcal{E}_i} \right. \nonumber \\
&\left. +V_{\text{eff}}(\zeta_{i,n},|\mathbf{r}_{n}-\mathbf{r}_{j}|)\frac{\partial C_{i,n}(\mathcal{E}_{i},|\mathbf{r}_{1}-\mathbf{r}_{i}|,...,|\mathbf{r}_{N_{\text{c}}}-\mathbf{r}_{i}|)}{\partial \mathcal{E}_i}\right]  \nonumber \\
&= {{\sum_{\substack{\;{i={N_{\text{c}}}+1} \\ {i} \neq {j}}}^{{N}}} \dot{\mathcal{E}_i}\sum_{n=1}^{N_{\text{c}}}c_{i,j}(t)C_{i,n}(\mathcal{E}_{i},|\mathbf{r}_{1}-\mathbf{r}_{i}|,...,|\mathbf{r}_{N_{\text{c}}}-\mathbf{r}_{i}|)\frac{\partial V_{\text{eff}}(\zeta_{i,n},|\mathbf{r}_{n}-\mathbf{r}_{j}|)}{\partial \zeta_{i,n}}\frac{\partial \zeta_{i,n}}{ \partial \mathcal{E}_i}  } \nonumber \\
&+ {{\sum_{\substack{\;{i={N_{\text{c}}}+1} \\ {i} \neq {j}}}^{{N}}} \dot{\mathcal{E}_i}\sum_{n=1}^{N_{\text{c}}}c_{i,j}(t)V_{\text{eff}}(\zeta_{i,n},|\mathbf{r}_{n}-\mathbf{r}_{j}|)\sum_{b=1}^{N_{\text{c}}}\left[ \frac{\partial C_{i,n}(\mathcal{E}_{i},|\mathbf{r}_{1}-\mathbf{r}_{i}|,...,|\mathbf{r}_{N_{\text{c}}}-\mathbf{r}_{i}|)}{\partial \zeta_{i,b}}\frac{\partial \zeta_{i,b}}{\partial \mathcal{E}_i} \right] } \nonumber\\
&= {{\sum_{\substack{\;{i={N_{\text{c}}}+1} \\ {i} \neq {j}}}^{{N}} g_{j,i} \dot{\mathcal{E}_i}  }}\\
&{\frac{\partial \mathcal{E}_{j}(t)}{\partial t}}  = {\sum_{\substack{\;{i={N_{\text{c}}}+1} \\ {i} \neq {j}}}^{{N}}}{\sum_{n=1}^{N_{\text{c}}}\dot{c}_{i,j}(t) C_{i,n}(\mathcal{E}_{i},|\mathbf{r}_{1}-\mathbf{r}_{i}|,...,|\mathbf{r}_{N_{\text{c}}}-\mathbf{r}_{i}|)V_{\text{eff}}(\zeta_{i,n},|\mathbf{r}_{n}-\mathbf{r}_{j}|) + \frac{\partial }{\partial t} \left\{ \frac{\left[\mathbf{\tilde{p}}_{{j}}- {Q_j}\mathbf{A}(\mathbf{r}_{{j}},{t}) \right]^2}{2{m_j}} \right\}  - Q_j\mathbf{r}_{j} \cdot \frac{\partial \mathbf{E}\left(\mathbf{r}_{j}, t\right)}{\partial t}} \nonumber \\
&= {\sum_{\substack{\;{i={N_{\text{c}}}+1} \\ {i} \neq {j}}}^{{N}}}{\sum_{n=1}^{N_{\text{c}}}\dot{c}_{i,j}(t) C_{i,n}(\mathcal{E}_{i},|\mathbf{r}_{1}-\mathbf{r}_{i}|,...,|\mathbf{r}_{N_{\text{c}}}-\mathbf{r}_{i}|)V_{\text{eff}}(\zeta_{i,n},|\mathbf{r}_{n}-\mathbf{r}_{j}|) + Q_j{\mathbf{\dot{r}}_{j} \cdot \mathbf{E}\left(\mathbf{r}_{j}, t\right) - Q_j\mathbf{r}_{j} \cdot \frac{\partial \mathbf{E}\left(\mathbf{r}_{j}, t\right)}{\partial t}}}
\end{align}
\end{widetext}}


\section{Derivatives of the functions of the effective charges. } \label{Appendix_B}
\subsection{Derivatives of $C_{i,n}(\mathcal{E}_{i},|\mathbf{r}_{1}-\mathbf{r}_{i}|,...,|\mathbf{r}_{N_{\text{c}}}-\mathbf{r}_{i}|)$}

The function $C_{i,n}(\mathcal{E}_{i},|\mathbf{r}_{1}-\mathbf{r}_{i}|,...,|\mathbf{r}_{N_{\text{c}}}-\mathbf{r}_{i}|)$ has the following derivative with respect to ${\zeta_{i,b}}$
\begin{align}\label{Cim_constant_derivative_zeta}
\begin{split}
&{\frac{\partial C_{i,n}(\mathcal{E}_{i},|\mathbf{r}_{1}-\mathbf{r}_{i}|,...,|\mathbf{r}_{N_{\text{c}}}-\mathbf{r}_{i}|)}{\partial \zeta_{i,b}}}\\
&=  {\frac{1}{\left( \sum_{{n'=1}}^{{N_{\text{c}}}}\rho_{i,n'} \right)^2} \frac{\partial \rho_{i,b}}{\partial \zeta_{i,b}}	\left(\delta_{n,b}\sum_{{n'=1}}^{{N_{\text{c}}}}\rho_{i,n'} -  \rho_{i,n} \right) }. 
\end{split}
\end{align}
The function $C_{i,b}(\mathcal{E}_{i},|\mathbf{r}_{1}-\mathbf{r}_{i}|,...,|\mathbf{r}_{N_{\text{c}}}-\mathbf{r}_{i}|)$ has the following derivative with respect to $\mathbf{r}_{n}$
\begin{align}\label{Cim_constant_derivative_rn}
\begin{split}
&{\frac{\partial C_{i,b}(\mathcal{E}_{i},|\mathbf{r}_{1}-\mathbf{r}_{i}|,...,|\mathbf{r}_{N_{\text{c}}}-\mathbf{r}_{i}|)}{\partial \mathbf{r}_{n}}} \\
&= {\frac{1}{\left( \sum_{{n'=1}}^{{N_{\text{c}}}}\rho_{i,n'} \right)^2} \frac{\partial \rho_{i,n}}{\partial \mathbf{r}_{n}}	\left(\delta_{n,b}\sum_{{n'=1}}^{{N_{\text{c}}}}\rho_{i,n'} -  \rho_{i,b} \right)}.
\end{split}
\end{align}
The function $C_{i,n}(\mathcal{E}_{i},|\mathbf{r}_{1}-\mathbf{r}_{i}|,...,|\mathbf{r}_{N_{\text{c}}}-\mathbf{r}_{i}|)$ has the following derivative with respect to $\mathbf{r}_{i}$
\begin{align}\label{Cim_constant_derivative_ri}
\begin{split}
&{\frac{\partial C_{i,n}(\mathcal{E}_{i},|\mathbf{r}_{1}-\mathbf{r}_{i}|,...,|\mathbf{r}_{N_{\text{c}}}-\mathbf{r}_{i}|)}{\partial \mathbf{r}_{i}}}\\
&= {\frac{\frac{\partial \rho_{i,n}}{\partial \mathbf{r}_{i}}  \left( \sum_{{n'=1}}^{{N_{\text{c}}}}\rho_{i,n'} \right) - \left( \sum_{{n'=1}}^{{N_{\text{c}}}}\frac{\partial  \rho_{i,n'}}{\partial \mathbf{r}_{i}} \right)\rho_{i,n}}{\left( \sum_{{n'=1}}^{{N_{\text{c}}}}\rho_{i,n'} \right)^2}  }.
\end{split}
\end{align}
The function $C_{j',n}(\mathcal{E}_{j'},|\mathbf{r}_{1}-\mathbf{r}_{j'}|,...,|\mathbf{r}_{N_{\text{c}}}-\mathbf{r}_{j'}|)$ has the following derivative with respect to $\mathbf{q}_k$, assuming that k can only represent electron-nucleus pairs
\begin{align}\label{Cim_constant_derivative_qk}
\begin{split}
&\frac{\partial C_{j',n}(\mathcal{E}_{j'},|\mathbf{r}_{1}-\mathbf{r}_{j'}|,...,|\mathbf{r}_{N_{\text{c}}}-\mathbf{r}_{j'}|)}{\partial \mathbf{q}_{k(i,j)}}=\\
%
%
 %
 %
%
%
&= \frac{\delta_{j',j}}{(\sum_{n' = 1}^{N_{\text{c}}} \rho_{j',n'})^2}\frac{\partial \rho_{j',i}}{\partial \mathbf{q}_{k}}\left(\delta_{n,i}  \sum_{n' = 1}^{N_{\text{c}}} \rho_{j',n'}  -{\rho_{j',n}  }\right) 
\end{split}
\end{align} 

The function ${\rho_{i,n}}$ has the following derivatives with respect to ${\mathbf{r}_{i}}$, ${\mathbf{r}_{n}}$, ${\mathbf{q}_{k}}$  and ${\zeta_{i,n}}$
{\allowdisplaybreaks
\begin{align}
\frac{\partial \rho_{i,n} }{\partial \mathbf{r}_{i}}   &=    2\zeta_{i,n}\frac{\mathbf{r}_n - \mathbf{r}_i}{|\mathbf{r}_n - \mathbf{r}_i|} \rho_{i,n} =  2\zeta_{i,n}\frac{\mathbf{q}_{k(n,i)}}{q_{k(n,i)}} \rho_{i,n}  \label{rho_constant_derivative_ri} \\
\frac{\partial \rho_{i,n} }{\partial \mathbf{r}_{n}}   &=   -2\zeta_{i,n}\frac{\mathbf{r}_n - \mathbf{r}_i}{|\mathbf{r}_n - \mathbf{r}_i|} \rho_{i,n} =  -2\zeta_{i,n}\frac{\mathbf{q}_{k(n,i)}}{q_{k(n,i)}} \rho_{i,n} \label{rho_constant_derivative_rn}\\
\frac{\partial \rho_{i,n} }{\partial \mathbf{q}_{k(i',j')}}   &=  \delta_{i',n}\delta_{j',i} \frac{\partial \rho_{i,n} }{\partial \mathbf{r}_{n}} \label{rho_constant_derivative_q} \\
\frac{\partial \rho_{i,n} }{\partial \zeta_{i,n}} &= \left(\frac{3}{\zeta_{i,n}} -2 q_{k(n,i)}\right) \rho_{i,n}. \label{rho_constant_derivative_zeta}
\end{align}
}

\subsection{Derivatives of $V_{\text{eff}}(\zeta_{i,n},|\mathbf{r}_{n}-\mathbf{r}_{j}|)$}
The function $V_{\text{eff}}(\zeta_{i,n},|\mathbf{r}_{n}-\mathbf{r}_{j}|)$ has the following derivative with respect to $\zeta_{i,n}$
\begin{align}\label{eq:Veff_derivative_zeta}
\begin{split}
\frac{\partial V_{\text{eff}}(\zeta_{i,n},|\mathbf{r}_{n}-\mathbf{r}_{j}|)}{\partial \zeta_{i,n}} &= e^{-2 \zeta_{i,n}|\mathbf{r}_{n}-\mathbf{r}_{j}| }\left(1 + 2 \zeta_{i,n} |\mathbf{r}_{n}-\mathbf{r}_{j}| \right). 
\end{split}
\end{align}
The function $V_{\text{eff}}(\zeta_{i,n},|\mathbf{r}_{n}-\mathbf{r}_{j}|)$ has the following derivative with respect to $\mathbf{r}_{n}$
\begin{widetext}
\begin{align}\label{eq:Veff_derivative_rn}
\begin{split}
&\dfrac{\partial V_{\text{eff}}(\zeta_{i,n},| \mathbf{r}_{n} - \mathbf{r}_{j} |)}{\partial \mathbf{r}_n}=\dfrac{-1+\left[ 1+2\zeta_{i,n} | \mathbf{r}_{n} - \mathbf{r}_{j} |(1+\zeta_{i,n} | \mathbf{r}_{n} - \mathbf{r}_{j} |)\right] e^{-2\zeta_{i,n} | \mathbf{r}_{n} - \mathbf{r}_{j} |}}{| \mathbf{r}_{n} - \mathbf{r}_{j} |^3} (\mathbf{r}_{n} - \mathbf{r}_{j}).
\end{split}
\end{align}
\end{widetext}
Finally, the function $V_{\text{eff}}(\zeta_{i,n},|\mathbf{r}_{n}-\mathbf{r}_{j}|)$ has the following derivative with respect to $\mathbf{q}_k$
\begin{align}\label{eq:Veff_derivative_qk}
\begin{split}
&\dfrac{\partial V_{\text{eff}}(\zeta_{j',i},| \mathbf{r}_{i} - \mathbf{r}_{i'} |)}{\partial \mathbf{q}_k}=\\
&\delta_{i',j}\dfrac{-1+\left[ 1+2\zeta_{j',i} q_{k}(1+\zeta_{j',i} q_k)\right] e^{-2\zeta_{j',i} q_k}}{q^3_k}\mathbf{q}_k.
\end{split}
\end{align}
\vspace{1cm}

\bibliography{bibliography}{}

\end{document}